\documentclass[11pt]{article}
\usepackage{amsmath,amssymb,cite,graphicx,slashed,color}
\usepackage[margin=1.2in]{geometry}

\def\Tr{\mathrm{Tr}}

\def\det{\mathrm{det}}
\def\calG{{\cal G}}
\def\calH{{\cal H}}
\def\calO{{\cal O}}
\def\calo{{\cal O}}
\def\calL{{\cal L}}
\def\GammaI{\Gamma^{\rm I}}
\def\GammaII{\Gamma^{\rm II}}
\def\tDelta{\widetilde\Delta}
\def\be{\begin{equation}}
\def\ee{\end{equation}}
\def\bea{\begin{eqnarray}}
\def\eea{\end{eqnarray}}
\def\eq#1{(\ref{#1})}

   {\setlength\paperheight {297mm}%
    \setlength\paperwidth  {210mm}}

\def\cG{{\cal G}}

\def\cO{{\cal O}}
\def\cR{{\cal R}}
\def\cC{{\cal C}}

\begin{document}

\centerline{\bf \large The Renormalization Group and Weyl--invariance}
\bigskip
\centerline{A. Codello, G. D'Odorico, C. Pagani, R. Percacci}
\medskip
\centerline{SISSA, via Bonomea 265, 34136 Trieste, Italy}
\centerline{and INFN, Sezione di Trieste, Italy}

\begin{abstract}
We consider matter fields conformally coupled 
to a background metric and dilaton and describe in detail a quantization procedure 
and related renormalization group flow that preserve Weyl invariance.
Even though the resulting effective action is Weyl--invariant, the trace anomaly is still present,
with all its physical consequences.
We discuss first the case of free matter and then extend the result to interacting matter.
We also consider the case when the metric and dilaton are dynamical and gravitons enter in the loops.
\end{abstract}


\section{Introduction}


The definition of a quantum field theory generally begins with a classical field theory 
with bare action $S$, which is then quantized by defining a functional integral.
Even if $S$ is scale- or (in curved spacetime) Weyl--invariant, the resulting quantum effective action in general is not, because in the definition of the functional integral one necessarily introduces a mass scale.
This is the origin of the celebrated trace anomaly \cite{duff}.

It has been known since early on that when a dilaton is present,
there is a way of perturbatively quantizing the theory which preserves Weyl invariance \cite{englert}.
This has been rediscovered several times in the literature \cite{fv,flop,reuterliouville,creh,machado,shapo}.
In this paper we will discuss mainly the implications of this type of quantization procedure for
the Renormalization Group (RG).
Actually, we will adopt a point of view that puts the RG first, and views the quantum effective action 
as the result of following the RG flow all the way to the IR.
We will use a nonperturbative, ``Wilsonian'' definition of the RG
which is seen as the dependence of the effective action on a cutoff that
is introduced by hand in the definition of the functional integral.
By using this definiton, we can extend the validity of the preceding statement
to any theory, independent of its renormalizability properties.

The discussion will be pedagogical and self--contained.
In the second section we discuss scale and Weyl invariance, and the notion of Weyl geometry.
We recall that when a dilaton is present, one can make any action Weyl invariant by 
replacing all dimensionful couplings by dimensionless couplings multiplied by powers of the dilaton.
This is a gravitational version of the ``St\"uckelberg trick''.
In the subsequent sections we show, in increasingly complicated cases,
that Weyl invariance can be preserved in the quantum theory.
We begin with massless, free matter fields conformally coupled to a background metric and dilaton. 
Since in this case the functional integral is Gaussian, one can prove directly 
that the effective action is Weyl--invariant.
In order to extend this statement to more complicated, interacting theories,
it turns out to be better to view the effective action as the IR endpoint of a Wilsonian RG flow.
One can then show quite generally that there exists a way of constructing the RG flow
which preserves Weyl--invariance, so if the initial point of the flow (in the UV) is Weyl--invariant,
also the IR endpoint will be.
For clarity we present this logic first in the case of free massless fields,
confirming the result obtained from direct evaluation of the effective action.
In these cases one can actually construct explicit one--parameter families of
Weyl--invariant actions that interpolate continuously between the bare action in the UV and
the effective action in the IR. The construction is complete in 2 dimensions,
where the effective action is (the Weyl--invariant version of) the Polyakov action, 
and limited to the first terms in a curvature expansion in 4 dimensions. 
We then observe that the condition of masslessness can be easily relaxed,
since a mass can be viewed as a coupling of the field to the dilaton.
Since any dimensionful coupling can be traded for a dimensionless coupling
times some power of the dilaton,
the proof can be extended to the case of interacting matter coupled to a background metric and dilaton
and finally to the case when gravity itself is quantized,
by which we mean that gravitons and dilatons are allowed to circulate in the loops.
In this way one has a fully nonperturbative proof that there exists a Weyl--invariant
definition of the effective action.

Normal physical theories are neither conformal-- nor scale--invariant.
The renormalization group running describes the dependence of couplings on
one dimensionful scale and the theory becomes conformally invariant only at a fixed point.
If we now reformulate an arbitrary theory in a Weyl--invariant way, several obvious questions arise:
What is the meaning of a cutoff in a Weyl--invariant theory?
What distinguishes a fixed point from any other point?
Are these Weyl--invariant quantum theories physically equivalent to
ordinary non--Weyl--invariant ones?
We will address these questions in the course of our derivations and
summarize the state of our understanding in the conclusions.


\section{Weyl invariance}


A global scale transformation is a rescaling of all lengths by a fixed, constant factor $\Omega$.
In flat space, scale transformations are usually interpreted as the map $x\to\Omega x$.
As such, they form a particular subgroup of diffeomorphisms.
Alternatively, one can think of rescaling the metric $g_{\mu\nu}\to\Omega^2 g_{\mu\nu}$.
The two points of view are completely equivalent, since lengths are
given by integrating the line element $ds=\sqrt{g_{\mu\nu}dx^\mu dx^\nu}$.
For our purposes it will be convenient to adopt the second point of view.

Let us now define the scaling dimension of a quantity.
Consider a theory with fields $\psi_a$, parameters $g_i$ (which include masses, couplings,
wave function renormalizations etc.) and action  $S(g_{\mu\nu},\psi_a,g_i)$.
There is a unique choice of numbers $w_a$ (one per field) and $w_i$ (one per parameter)
such that $S$ is invariant: 
\be
\label{weylinv}
S(g_{\mu\nu},\psi_a,g_i)=S(\Omega^2 g_{\mu\nu},\Omega^{w_a}\psi_a,\Omega^{w_i}g_i)\ .
\ee
(It does not matter here whether the metric is fixed or dynamical.)
The numbers $w_a$, $w_i$ are called the scaling dimensions, or the weights, of $\psi_a$ and $g_i$.
In this paper we will assume that the spacetime coordinates are dimensionless
and we use natural units where $c=1$, $\hbar=1$.
Then, the scaling dimensions are equal to the ordinary length dimensions of $\psi_a$ and $g_i$
in the sense of dimensional analysis.
Since in particle physics it is customary to use mass dimensions,
when we talk of ``dimensions'' without further specification we will refer to
the mass dimensions $d_a=-w_a$ and $d_i=-w_i$ .
In $d$ spacetime dimensions, the dimensions of scalar, spinor and vector fields 
are $(d-2)/2$, $(d-1)/2$ and $(d-4)/2$, respectively. 
One can easily convince oneself that the dimensions of all parameters in the Lagrangian, 
such as masses and couplings, are the same as in the more familiar case when coordinates have 
dimension of length.

Changing couplings is usually interpreted as changing theory,
so in general the transformations \eq{weylinv} are not symmetries
of a theory but rather maps from one theory to another.
In the case when all the $w_i$ are equal to zero, we have
\be
\label{scaleinv}
S(g_{\mu\nu},\psi_a,g_i)=S(\Omega^2 g_{\mu\nu},\Omega^{w_a}\psi_a,g_i)\ .
\ee
Since these are transformations that map a theory to itself,
a theory of this type is said to be globally scale invariant.

Scale transformations with $\Omega$ a positive real function of $x$
are called Weyl transformations.
They act on the metric and the fields exactly as in \eq{weylinv}.
What about the parameters? They are supposed to be $x$-independent,
so transformation $g_i\to \Omega(x)^{w_i}g_i$ would not make much sense.
One can overcome this difficulty by promoting the dimensionful parameters to fields.
One can then meaningfully ask whether \eq{weylinv} holds.
In general the answer will be negative, but there is a simple procedure
that allows one to make a scale invariant theory also Weyl--invariant:
it is called Weyl gauging and it was the earliest incarnation of the
notion of gauge theory.
In this paper we will restrict ourselves to a special case of Weyl gauging, namely the case
when the connection is flat.
We pick a mass parameter of the theory, let's call it $\mu$ and we promote 
it to a function that we shall denote $\chi$. We can write 
\be
\label{sigma}
\chi(x)=\mu e^{\sigma(x)}\ ,
\ee
where $\mu$ is constant.
The function $\chi$, or sometimes $\sigma$, is called the dilaton.
Notice that unlike an ordinary scalar field, it has dimension one independently of
the spacetime dimensionality. Thus it transforms under Weyl transformation
as $\chi\mapsto\Omega^{-1}\chi$.
Now we can take any other dimensionful coupling of the theory and write
\be
\label{dimlesscouplings}
g_i=\chi^{-w_i}\hat g_i=\chi^{d_i}\hat g_i\ ,
\ee
where $\hat g_i$ is dimensionless (and therefore Weyl--invariant).
In general, a caret over a symbol denotes the same quantity measured in units of the dilaton. 
In principle one could promote more than one dimensionful parameter,
or even all dimensionful parameters, to independent dilatons.
This may have interesting applications, but for the sake of
simplicity in this paper we shall restrict ourselves to the case
when there is a single dilaton.

With the dilaton we construct a pure-gauge abelian gauge field
$b_\mu=-\chi^{-1}\partial_\mu\chi$, transforming under \eq{weylinv} as
$b_\mu\mapsto b_\mu+\Omega^{-1}\partial_\mu\Omega$.
Let $\nabla_\mu$ be the covariant derivative with respect to the
Levi-Civita connection of the metric $g$.
Define a new (non-metric) connection
\be
\label{hatGamma}
\hat\Gamma_\mu{}^\lambda{}_\nu=
\Gamma_\mu{}^\lambda{}_\nu-\delta^\lambda_\mu b_\nu-\delta^\lambda_\nu b_\mu
+g_{\mu\nu}b^\lambda\ ,
\ee
where $\Gamma_\mu{}^\lambda{}_\nu$ are the Christoffel symbols of $g$.
The corresponding covariant derivative is denoted $\hat\nabla$.
The connection coefficients $\hat\Gamma$ are invariant under \eq{weylinv}.
For any tensor $t$ of weight $w$ define the covariant derivative $Dt$ to be
\be
\label{wcovder}
D_\mu t=\hat\nabla_\mu t-wb_\mu t\ ,
\ee
where all indices have been suppressed.
We see that the weigth (or the dimension) acts like the Weyl charge of the field.
The tensor $Dt$ is covariant under diffeomorphisms and under Weyl transformations.
The curvature of $D$ is defined by
\be
[D_\mu,D_\nu]v^\rho=\cR_{\mu\nu}{}^\rho{}_\sigma v^\sigma\ .
\ee
The tensor $\cR_{\mu\nu}{}^\rho{}_\sigma$ is Weyl invariant,
and raising and lowering indices one obtains Weyl covariant expressions
of different dimensions.
A direct calculation gives the explicit expression
\bea
\label{wcovcurv}
\cR_{\mu\nu\rho\sigma}&=& R_{\mu\nu\rho\sigma}
+g_{\mu\rho}\left(\nabla_\nu b_\sigma+b_\nu b_\sigma\right)
-g_{\mu\sigma}\left(\nabla_\nu b_\rho+b_\nu b_\rho\right)
\nonumber
\\
&&
\!\!\!\!\!\!\!\!\!\!\!\!\!\!\!\!\!\!
-g_{\nu\rho}\left(\nabla_\mu b_\sigma+b_\mu b_\sigma\right)
+g_{\nu\sigma}\left(\nabla_\mu b_\rho+b_\mu b_\rho\right)
-\left(g_{\mu\rho}g_{\nu\sigma}-g_{\mu\sigma}g_{\nu\rho}\right)b^2\ .
\eea
From here one finds the analogs of the Ricci tensor and Ricci scalar
\bea
\cR_{\mu\nu}&=&R_{\mu\nu}+(d-2)b_\mu b_\nu+(d-2)\nabla_\mu b_\nu
-(d-2)b^2g_{\mu\nu}+\nabla^\rho b_\rho g_{\mu\nu}\ ,
\\
\cR&=&R+2(d-1)\nabla^\mu b_{\mu}-(d-1)(d-2)b^2\ .
\eea
It is also possible to define the tensor $\cC^\mu{}_{\nu\alpha\beta}$
which is related to $\cR^\mu{}_{\nu\alpha\beta}$ by the same formula that
relates $C^\mu{}_{\nu\alpha\beta}$ to $R^\mu{}_{\nu\alpha\beta}$, and therefore
reduces to the standard Weyl tensor in a gauge where $\chi$ is constant.

Now start from a generic action for matter and gravity of the form
$S(g_{\mu\nu},\psi_a,g_i)$.
Express every parameter $g_i$ as in \eq{dimlesscouplings}.
Replace all covariant derivatives $\nabla$ by Weyl covariant derivatives $D$ and all curvatures $R$ by the Weyl covariant curvatures $\cR$.
Now all the terms appearing in the action are products of Weyl covariant
objects, and local Weyl invariance just follows from the fact that the action
is dimensionless.
In this way we have defined an action $\hat S(g_{\mu\nu},\chi,\psi_a,\hat g_i)$.
It contains only dimensionless couplings $\hat g_i$, 
and is Weyl invariant by construction.
One can choose a gauge where $\chi=\mu$ is constant (equivalently, $\sigma=0$),
and in this gauge the action $\hat S(g_{\mu\nu},\chi,\psi_a,\hat g_i)$ 
reduces to the original one.

The above construction defines an ``integrable Weyl geometry'',
since the curvature of the Weyl gauge field $b_\mu$ is zero.
In this integrable case there is also another way of defining a Weyl--invariant action
from a non--invariant one, namely to replace all the arguments in $S$ by the corresponding
dimensionless quantities $\hat g_{\mu\nu}=\chi^2g_{\mu\nu}$,
$\hat\psi_a=\chi^{w_a}\psi_a$ and $\hat g_i=\chi^{w_i}g_i$
and subsequently reexpress the action in terms of the original fields
\be
\hat S(g_{\mu\nu},\chi,\psi_a,\hat g_i)=
S(\hat g_{\mu\nu},\hat\psi_a,\hat g_i)\ .
\ee
It is easy to see that this construction gives the same result as the preceding one.
This follows from the fact that \eq{hatGamma} are the Christoffel symbols of $\hat g_{\mu\nu}$,
that $\hat\nabla_\mu\hat\psi_a=\chi^{w_a}D_\mu\psi_a$ and that the curvature tensor
of $\hat\Gamma$ is $\cR_{\mu\nu\rho\sigma}$.
\footnote{If we call $\hat R_{\mu\nu\rho\sigma}$ the Riemann tensor of $\hat g_{\mu\nu}$,
we have $\hat R^\mu{}_{\nu\rho\sigma}=\cR^\mu{}_{\nu\rho\sigma}$
and $\hat R_{\mu\nu\rho\sigma}=\chi^2\cR_{\mu\nu\rho\sigma}$.}

The above procedure can be used to rewrite any theory in Weyl--invariant form.
Not all Weyl--invariant theories are of this type: there are also theories
that are Weyl--invariant without containing a Weyl gauge field $b_\mu$
(or a dilaton). In such theories the terms generated by a Weyl transformation
that contain the derivatives of the transformation parameter are compensated by terms
generated by variations of Ricci tensors. Since Weyl--invariance can be viewed as a
gauged version of global scale invariance, this has been called ``Ricci gauging'' in \cite{iorio}.
It was also shown that such Ricci--gauged theories correspond (under mild additional assumptions)
to theories that are conformal--invariant, as opposed to merely scale--invariant, in flat space.
The existence of well--behaved theories that are scale-- but not conformal--invariant in flat space has been reexamined recently \cite{jackiw,fortin}.

\section{The effective action of free matter fields coupled to an external gravitational field}

\subsection{The standard measure}

In this section we review the evaluation of the effective action for free,
massless matter fields conformally coupled to a metric.
This will provide the basis for
different quantization procedures to be described in the following.
Much of the discussion can be carried out in arbitrary even dimension $d$.

For definiteness let us consider first a single conformally coupled scalar field,
with equation of motion $\Delta^{(0)}\phi=0$, where
$\Delta^{(0)}=-\nabla^2+\frac{d-2}{4(d-1)}R$.
Functional integration over $\phi$ in the presence of a source $j$
leads to a generating functional $W(g_{\mu\nu},j)$, 
whose Legendre transform $\Gamma(g_{\mu\nu},\phi)=W(g_{\mu\nu},j)-\int j\phi$
is the effective action.
For the definition of the functional integral one needs a metric 
(more precisely an inner product) in the space of the fields. We choose
\be
\cG(\phi,\phi')=\mu^2\int dx\sqrt{g}\,\phi\,\phi'\ ,
\ee
where $\mu$ is an arbitrary mass that has to be introduced for dimensional reasons.
The action can be written as
\be
\label{gaussianaction}
S_S(g_{\mu\nu},\phi)=
\frac{1}{2}\int dx\sqrt{g}\,\phi \Delta^{(0)}\phi
=\frac{1}{2}\cG\!\left(\phi,\frac{\Delta^{(0)}}{\mu^2}\phi\right)
=\frac{1}{2}\sum_n a_n^2\lambda_n/\mu^2 \ ,
\ee
where $\lambda_n$ are the eigenvalues of $\Delta^{(0)}$, $\phi_n$ the corresponding eigenfunctions
and $a_n$ are the (dimensionless) coefficients of the expansion of $\phi$
on the basis of the eigenfunctions:
\be
\label{basis}
\Delta^{(0)}\phi_n=\lambda_n\phi_n\ ;\qquad
\cG(\phi_n,\phi_m)=\delta_{nm}\ ;\qquad
\phi=\sum_n a_n\phi_n\ ;\qquad
a_n=\cG(\phi,\phi_n)\ .
\ee
(For simplicity we assume that the manifold is compact and without boundary,
so that the spectrum of the Laplacian is discrete.)
Weyl--covariance means that under a Weyl transformation 
the operator $\Delta^{(0)}$ transforms as
\be
\label{optransf}
\Delta^{(0)}_{\Omega^2 g}=\Omega^{-1-\frac{d}{2}}\Delta^{(0)}_g\Omega^{\frac{d}{2}-1}\ ,
\ee
where we have made the dependence of the metric explicit.
For an infinitesimal transformation $\Omega=1+\omega$,
\be
\label{optransfinf}
\delta_\omega\Delta^{(0)}=-2\omega\Delta^{(0)}\ .
\ee
The functional measure is $(d\phi)=\prod_n da_n$,
so the Gaussian integral can be evaluated as
\be
e^{-W(g_{\mu\nu},j)}=
\prod_n\left(\int da_n e^{-\frac{1}{2}a_n^2 \lambda_n/\mu^2-a_n j^n}\right)=
\prod_n\sqrt{\frac{\mu}{\lambda_n}}e^{\frac{1}{2}\frac{\mu^2}{\lambda_n}(j^n)^2}=
\det\left(\frac{\Delta^{(0)}}{\mu^2}\right)^{-1/2}e^{\frac{1}{2}\int j\Delta^{-1}j}
\ee
up to a field--independent multiplicative constant. 
From here one gets (using the same notation for the VEV as for the field)
$\phi=-\Delta^{(0)-1}j$, so finally the Legendre transform gives
\be
\label{oneloop}
\Gamma(\phi,g_{\mu\nu})=S_S(\phi,g_{\mu\nu})+\frac{1}{2}\Tr\log\left(\frac{\Delta^{(0)}}{\mu^2}\right)\ .
\ee
An UV regularization is needed to define this trace properly.
We see that the scale $\mu$, which has been introduced in the definition
of the measure, has made its way into the functional determinant.

Things work much in the same way for the fermion field,
which contributes to the effective action a term
\be
S_D(\bar\psi,\psi,g_{\mu\nu})-\frac{1}{2}\Tr\log\left(\frac{\Delta^{(1/2)}}{\mu^2}\right)\ ,
\ee
where $S_D$ is the classical action and
$\Delta^{(1/2)}=-\nabla^2+\frac{R}{4}$ is the square of the Dirac operator.

The Maxwell action is Weyl--invariant only in $d=4$.
With our conventions the field $A_\mu$ is dimensionless
and the Weyl--invariant inner product in field space is:
\be
\label{prodmax}
\cG(A,A')=\mu^2\int d^4x\sqrt{g}\,g^{\mu\nu}A_\mu A_\nu\ .
\ee
Using the standard Faddeev-Popov procedure, we add gauge fixing and ghost actions
\be
S_{GF}=\frac{1}{2\alpha}\int d^4x\sqrt{g}\,(\nabla_\mu A^\mu)^2\ ;\qquad
S_{gh}=\int d^4x\sqrt{g}\,\bar C\Delta^{(gh)}C\ ,
\ee
with $\Delta^{(gh)}=-\nabla^2$.
Then, in the gauge $\alpha=1$, the gauge--fixed action becomes
\be
S_M+S_{GF}=\frac{1}{2}\int d^4x\sqrt{g}\,A^\mu \Delta_\mu^{(1)\nu} A_\nu
=\frac{1}{2}\cG\left(A,\frac{\Delta^{(1)}}{\mu^2}A\right)\ ,
\ee
where $\Delta_\mu^{(1)\nu}=-\nabla^2\delta_\mu^\nu+R_\mu^\nu$ is the Laplacian on one--forms.
Following the same steps as for the scalar field, we obtain a contribution to the
effective action equal to
\be
\label{oneloopmaxwell}
S_M(A_\mu,g_{\mu\nu})+
\frac{1}{2}\Tr\log\left(\frac{\Delta^{(1)}}{\mu^2}\right)
-\Tr\log\left(\frac{\Delta^{(gh)}}{\mu^2}\right)\ .
\ee
Note that even though the Maxwell action $S_M$ is Weyl--invariant,
the gauge fixing action is not, nor is the ghost action.
As a result the operators $\Delta^{(1)}$ and $\Delta^{(gh)}$ are not Weyl--covariant.
Instead of an equation like (\ref{optransfinf}), they satisfy (in four dimensions)
\bea
\delta_\omega\Delta^{(gh)}&=&-2\omega\Delta^{(h)}-2\nabla^\nu\omega\nabla_\nu\ ;
\\
\delta_\omega\Delta^{(1)}_\mu{}^\nu&=&-2\omega\Delta^{(1)}_\mu{}^\nu
+2\nabla_\mu\omega\nabla^\nu-2\nabla^\nu\omega\nabla_\mu-2\nabla_\mu\nabla^\nu\omega\ .
\eea
We shall see in the next section how these non--invariances compensate each other in the effective action,
so that the breaking of Weyl--invariance is only due to the presence of the scale $\mu$
which was introduced in the inner product.

In general, the need for an inner product in field space can also be seen 
in a more geometrical way as follows.
The classical action, being quadratic in the fields, has the form 
$\calH(\phi,\phi)$, where $\calH=\frac{\delta^2S}{\delta\phi\delta\phi}$
can be viewed as a covariant symmetric tensor in field space:
when contracted with a field (a vector in field space)
it produces a one--form in field space.
Now, the determinant of a covariant symmetric tensor is not a basis-independent
quantity. One can only define in a basis-independent way the determinant
of an operator mapping a space into itself, i.e. a mixed tensor.
One can transform the covariant tensor $\calH$ to a mixed tensor $\calO$
by ``raising an index'' with a metric:
\footnote{In de Witt's condensed notation, where an index $i$ stands
both for a point $x$ in spacetime and whatever tensor or spinor indices
the field may be carrying, this equation reads $\calO_i{}^j=\calH_{ik}\calG^{kj}$.}
\be
\label{raise}
\calH(\phi,\phi')=\cG(\phi,\calO\phi')\ .
\ee
It is the determinant of the operator $\calO$ that appears in the effective action.
Again we see that the scale $\mu$ appears through the metric $\calG$,
which is needed to define the determinant.
Notice that since $\calO\phi$ is another field of the same type as $\phi$,
$\calO$ must necessarily be dimensionless, and this is guaranteed by the factors of $\mu$
contained in $\calG$. For example, in the scalar case,
$\calO=\frac{1}{\mu^2}\Delta^{(0)}$.

\subsection{Trace anomaly}

Under an infinitesimal Weyl transformation 
the variation of the effective action is
\be
\delta_\omega\Gamma=
\int dx\,\frac{\delta\Gamma}{\delta g_{\mu\nu}}2\omega g_{\mu\nu}
=-\int dx\sqrt{g}\,\omega\langle T^\mu_\mu\rangle\ .
\ee
The trace of the energy--momentum tensor vanishes for a Weyl--invariant action,
so the appearance of a nonzero trace is the physical manifestation of the anomaly.

For non--interacting fields the one--loop effective action is exact.
From the proper time representation
\be
\label{propertime}
\Gamma=S-\frac{1}{2}\int_{\epsilon/\mu^2}^\infty \frac{dt}{t}\Tr e^{-t\Delta}\ ,
\ee
where $\epsilon$ is a dimensionless UV regulator,
and from the Weyl covariance $\delta_\omega\Delta=-2\omega\Delta$
one finds
$$
\delta_\omega\Gamma
=\frac{1}{2}\int_{\epsilon/\mu^2}^\infty\!\!\! dt\,\Tr\,\delta_\omega\Delta e^{-t\Delta}
=-\!\!\int_{\epsilon/\mu^2}^\infty\!\! dt\,\Tr(\omega\Delta e^{-t\Delta})
=\int_{\epsilon/\mu^2}^\infty\!\! dt\frac{d}{dt}\Tr(\omega e^{-t\Delta})
=-\Tr\left[\omega e^{-\epsilon\Delta/\mu^2}\right]\ .
$$
For $\epsilon\to0$ one has from the asymptotic expansion of the heat kernel:
\be
\label{hk}
\Tr\left[\omega e^{-\epsilon\Delta/\mu^2}\right]=
\frac{1}{\left(4\pi\right)^{d/2}}\int d^dx\sqrt{g}\,\omega
\Bigl[
\frac{\mu^d}{\epsilon^{d/2}} b_{0}({\Delta})
+\frac{\mu^{d-2}}{\epsilon^{d/2-1}}b_{2}({\Delta})+\ldots
+b_d(\Delta)+\ldots\,\Bigr]\ ,
\ee
where $b_i$ are scalars constructed with $i$ derivatives of the metric.
All terms $b_i$ with $i>d$ tend to zero in the limit, 
so assuming that the power divergences (for $i<d$) are removed by renormalization,
there remains a universal, finite limit
\be
\delta_\omega\Gamma=-\frac{1}{\left(4\pi\right)^{d/2}}\int dx\sqrt{g}\,\omega\, b_d(\Delta)\ .
\ee
which implies that 
\be
\label{b4}
\langle T^\mu_\mu\rangle=b_d(\Delta)\ .
\ee
We note that this can also be seen as a direct manifestation of the dependence of
the result on the scale $\mu$. In fact one has, formally
\be
\label{tb}
\mu\frac{d}{d\mu}\frac{1}{2}\Tr\log\frac{\Delta}{\mu^2}=-\Tr\mathbf{1}
=-\int dx\sqrt{g}\,b_d(\Delta)
=-\int dx\sqrt{g}\,\langle T^\mu_\mu\rangle\ ,
\ee
where in the second step we have used zeta function regularization \cite{hawking}.

Aside from the different prefactor the calculation follows the same steps
in the case of massless spinors.
The Maxwell field, however, requires some additional considerations,
because the operators $\Delta^{(1)}$ and $\Delta^{(gh)}$ that appear in
\eqref{oneloopmaxwell} are not covariant. (We restrict ourselves now to $d=4$).
The first two steps of the preceding calculation give:
\bea
\delta_\omega\Gamma
&=&\frac{1}{2}\int_{\epsilon/\mu^2}^\infty\!\!\! dt\,\Tr\,\delta_\omega\Delta^{(1)} e^{-t\Delta^{(1)}}
-\int_{\epsilon/\mu^2}^\infty\!\!\! dt\,\Tr\,\delta_\omega\Delta^{(gh)} e^{-t\Delta^{(gh)}}
\nonumber
\\
\label{jules}
&=&\frac{1}{2}\int_{\epsilon/\mu^2}^\infty\!\!\! dt\,\Tr\,
(-2\omega\Delta^{(1)}+\rho^{(1)})
 e^{-t\Delta^{(1)}}
-\int_{\epsilon/\mu^2}^\infty\!\!\! dt\,\Tr\,
(-2\omega\Delta^{(gh)}+\rho^{(gh)})
e^{-t\Delta^{(gh)}}
\eea
where the violation of Weyl covariance is due to
\be
\rho^{(gh)}=-2\nabla^\nu\omega\nabla_\nu\ ;
\qquad
\rho^{(1)}_\mu{}^\nu=2\nabla_\mu\omega\nabla^\nu-2\nabla^\nu\omega\nabla_\mu-2\nabla_\mu\nabla^\nu\omega\ .
\ee
Since $\Delta^{(1)}$ maps longitudinal fields to longitudinal fields
and transverse fields to transverse fields, $\rho^{(1)}e^{-t\Delta^{(1)}}$ has vanishing matrix elements 
between transverse gauge fields.
Therefore the trace containing $\rho^{(1)}$ can be restricted to the subspace of longitudinal gauge potentials.
Let $\phi_n$ be a basis of eigenfunctions of $\Delta^{(gh)}$ satisfying an orthonormality
condition as in \eqref{basis}.
Then a basis in the space of longitudinal potentials satisfying a similar
orthonormality condition with respect to the inner product \eqref{prodmax}
is given by the fields $A^L_{n\mu}=\frac{1}{\sqrt{\lambda_n}}\nabla_\mu\phi_n$.
The traces of the terms violating Weyl--covariance are therefore:
\be
\label{peter}
\frac{1}{2}\mathrm{Tr}\rho^{(1)}e^{-t\Delta^{(1)}}-\mathrm{Tr}\rho^{(gh)}e^{-t\Delta^{(gh)}}
=\frac{1}{2}\sum_n\cG\left(A^L_n,\rho^{(1)}e^{-t\Delta^{(1)}}A^L_n\right)
-\sum_n\cG\left(\phi_n,\rho^{(gh)}e^{-t\Delta^{(gh)}}\phi_n\right) \ .
\ee
Noting that
$$
\Delta^{(1)}A^L_n=\frac{1}{\sqrt{\lambda_n}}\Delta^{(1)}\nabla_\mu\phi_n
=\frac{1}{\sqrt{\lambda_n}}\nabla_\mu\Delta^{(gh)}\phi_n
=\lambda_nA^L_n\ ,
$$
we can evaluate the matrix elements: 
$$
\cG\left(A^L_n,\rho^{(1)}e^{-t\Delta^{(1)}}A^L_n\right) 
=-4e^{-t\lambda_n}\cG\left(\phi_n,\nabla^\nu\omega\nabla_\nu\phi_n
\right)\ ,
$$
whereas in the ghost trace we have
$$
\cG\left(\phi_n,\rho^{(gh)}e^{-t\Delta^{(gh)}}\phi_n\right) 
=-2e^{-t\lambda_n}\cG\left(\phi_n,\nabla^\nu\omega\nabla_\nu\phi_n\right)\ .
$$
We see that the sums in \eqref{peter} cancel mode by mode.
As a result only the first term remains in each of the traces in \eqref{jules}.
From this point onwards the calculation proceeds as in the case of the scalar
and finally gives
\be
\delta_\omega\Gamma=\frac{1}{(4\pi)^2}\int d^4x\sqrt{g}
\left[b_4(\Delta^{(1)})-2b_4(\Delta^{(gh)})\right]\ .
\ee

The coefficients of the expansion of the heat kernel for Laplace-type
operators are well-known.
If there are $n_S$ scalar, $n_D$ spinors, one has in two dimensions
\be
\label{traceanomaly2}
\langle T^\mu{}_\mu\rangle=\frac{c}{24\pi} R
\ee
with
\be
c=n_S+n_D
\ee
whereas in four dimensions (assuming also the existence of $n_M$ Maxwell fields)
\be
\label{traceanomaly4}
\langle T^\mu{}_\mu\rangle
=c\,C^2-aE
\ee
where
$E=R_{\mu\nu\rho\sigma}R^{\mu\nu\rho\sigma}-4R_{\mu\nu}R^{\mu\nu}+R^2$
is the integrand of the Euler invariant,
$C^2=C_{\mu\nu\rho\sigma}C^{\mu\nu\rho\sigma}$
is the square of the Weyl tensor and the anomaly coefficients are
\footnote{the coefficients $c$ and $a$ were called $b$ and $-b'$ in \cite{duff}.}
\be
a={1\over360(4\pi)^2}\left(n_S+11n_D+62n_M\right)\ ;\quad
c={1\over120(4\pi)^2}\left(n_S+6n_D+12n_M\right)\ .
\ee

\subsection{The Weyl--invariant measure}

Let us now assume that the theory contains also a dilaton $\chi$.
For the purposes of this section it will be considered as part of
the gravitational sector and treated as an external field.
For notational simplicity we will discuss the case $d=4$ but it is easy
to generalize to arbitrary even dimensions.

The crucial observation is that we can now construct Weyl invariant metrics
in the spaces of scalar, Dirac and Maxwell fields, replacing the fixed scale $\mu$ by the dilaton:
\bea
{\cal G}_S(\phi,\phi')&=&\int d^{4}x\sqrt{g}\,\chi^{2}\phi\phi'\ ,\\
{\cal G}_D(\bar{\psi},\psi')&=&\int d^{4}x\sqrt{g}\,\frac{1}{2}\chi[\bar{\psi}\psi'+\bar{\psi}'\psi]\ ,\\
{\cal G}_M(A,A')&=&\int d^{4}x\sqrt{g}\,\chi^{2}A_{\mu}g^{\mu\nu}A'_{\nu}\ .
\label{wimM}
\eea
One can follow step by step the calculation in section 3.2,
the only change being the replacement of $\mu$ by $\chi$.
The final result for the one--loop contribution to the effective action can be written as
\be
\frac{n_S}{2}\Tr\log{\cal O}_S
-\frac{n_D}{2}\Tr\log{\cal O}_D
+\frac{n_M}{2}\Tr\log{\cal O}_M-n_M\Tr\log{\cal O}_{gh}\ ,
\ee
where now
\begin{align}
\calO_S&=\chi^{-2}\Delta^{(0)}\ ,\\
\calO_D&=\chi^{-2}\Delta^{(1/2)}\ ,\\
\calO_{M\mu}{}^{\nu}&=\chi^{-2}g_{\mu\sigma}\left(\Delta^{(1)}\right)^{\sigma\nu}\ ,\\
\calO_{gh}&=\chi^{-2}\Delta^{(gh)}\ ,
\end{align}
One can then verify that
\begin{align}
\calO_S^{\Omega}(\Omega^{-1}\phi)&=\Omega^{-1}{\cal O}_S\phi
\\
\calO_D^{\Omega}(\Omega^{-3/2}\psi)&=\Omega^{-3/2}\calO_D\psi
\\
\calO^{\Omega}_M{}_{\mu}{}^{\nu}A_{\nu}&=\calO_{M\mu}{}^{\nu}A_{\nu}
\\
{\cal O}_{gh}^{\Omega}(\Omega^{-1}c)&=\Omega^{-1}\calO_{gh}c.
\end{align}
where the notation $\calO^\Omega$ stands for the operator $\calO$ constructed with the
transformed metric $g^\Omega=\Omega^2g$ and dilaton $\chi^\Omega=\Omega^{-1}\chi$.
These operators map fields into fields transforming in the same way.
(As observed earlier, they are dimensionless.)
This implies that the eigenvalues of the operators $\calO$ are Weyl--invariant and
therefore also their determinants are invariant.
We conclude that in the presence of a dilaton there exists a
quantization procedure for noninteracting matter fields that respects Weyl invariance.

\subsection{The Wess--Zumino action}

We have seen that in the presence of a dilaton one has a choice between
different quantization procedures, which can be understood as different functional measures:
one of them breaks Weyl--invariance while the other maintains it.
Let us denote $\GammaI$ the effective action obtained with the standard measure 
and $\GammaII$ the one obtained with the Weyl--invariant measure.
The first is anomalous:
\be
\delta_\omega\GammaI=\int dx\,2\omega\frac{\delta\GammaI}{\delta g_{\mu\nu}}g_{\mu\nu}
=-\int dx\sqrt{g}\,\omega\langle T^\mu{}_\mu\rangle^I\not=0
\ee
whereas the second is Weyl invariant: $\GammaII(g^\Omega,\chi^\Omega)=\GammaII(g,\chi)$,
or in infinitesimal form
\be
\label{olga}
0=\delta_\omega\GammaII=\int dx\sqrt{g}\,\omega
\left(2\frac{\delta\GammaII}{\delta g_{\mu\nu}}g_{\mu\nu}-
\frac{\delta\GammaII}{\delta\chi}\chi\right)\ .
\ee
The Weyl invariant measure differs from the standard one simply by the
replacement of the fixed mass $\mu$ by the dilaton $\chi$, therefore we have
\be
\GammaII(g_{\mu\nu},\mu)=\GammaI(g_{\mu\nu})\ .
\ee
We see that $\GammaII$ can be obtained from $\GammaI$ by applying the
St\"uckelberg trick {\it after} quantization, {\it i.e.} to the mass parameter
$\mu$ that has been introduced by the functional measure.

Another useful point of view is the following.
Noting that $\Omega=\chi/\mu$ can be interpreted as the parameter of a Weyl transformation,
the variation of $\GammaI$ under a finite Weyl transformation 
defines a functional $\Gamma_{WZ}(g,\chi)$, the so-called ``Wess-Zumino action'', by:
\footnote{Here we view the Wess-Zumino action as a functional of a metric and a dilaton,
two dimensionful fields. Sometimes one may prefer to think of it as as a functional
of a metric and a Weyl transformation, the latter being a dimensionless function.
The two points of view are related by some factors of $\mu$.}
\be
\label{wz}
\GammaI(g^\Omega)-\GammaI(g)=\Gamma_{WZ}(g,\mu\Omega)
\ee
It satisfies the so-called Wess-Zumino consistency condition,
which can be written in the form
\be
\label{wzconsistency}
\Gamma_{WZ}(g^\Omega,\chi^\Omega)-\Gamma_{WZ}(g,\chi)=-\Gamma_{WZ}(g,\mu\Omega)
\ee
where $g^\Omega=\Omega^2 g$, $\chi^\Omega=\Omega^{-1}\chi$.
This shows that variation of the WZ action under a Weyl transformation
is exactly opposite to that of the action $\GammaI$.
From these definitions we see that the $\chi$-dependence of the Weyl--invariant action
is entirely contained in a Wess-Zumino term:
\be
\GammaII(g,\chi)=\GammaI(g)+\Gamma_{WZ}(g,\chi)\ .
\ee
We can think of the Weyl--invariant effective action as the ordinary effective
action to which a Wess-Zumino term has been added, with the effect of canceling the Weyl anomaly.
\footnote{This is completely analogous to what happens with gauge invariance in chiral theories
\cite{raja}.}

In the case of non--interacting, massless, conformal matter fields
the WZ actions can be computed explicitly by integrating the trace anomaly.
Let $\Omega_t$ be a one-parameter family of Weyl transformations 
with $\Omega_0=1$ and $\Omega_1=\Omega$, and let $g(t)_{\mu\nu}=g^{\Omega(t)}_{\mu\nu}$.
\be
\Gamma_{WZ}(g_{\mu\nu},\Omega)
=\int_0^1 dt \int dx\,\frac{\delta\Gamma}{\delta g_{\mu\nu}}\Bigg|_{g(t)}\delta g(t)_{\mu\nu}
=-\int_0^1 dt \int dx\,\sqrt{g(t)}\langle T^\mu_\mu\rangle_k\Omega(t)^{-1}\frac{d\Omega}{dt}\ .
\ee
In two dimensions, integrating the anomaly (\ref{traceanomaly2}) 
and using the parametrization (\ref{sigma}), one finds
\be
\label{wz2}
\Gamma_{WZ}(g_{\mu\nu},\mu e^\sigma)=-\frac{c}{24\pi}\int d^2x\sqrt{g}\left(R\sigma-\sigma\nabla^2\sigma\right)\ .
\ee
A similar procedure in four dimensions using (\ref{traceanomaly4}) leads to
\begin{equation}
\label{wz4}
\Gamma_{WZ}(g_{\mu\nu},\mu e^\sigma)
=-\int dx\,\sqrt{g}\left\{
c\,C^2\sigma-a\left[\left(E-\frac{2}{3}\Box R\right)\sigma+2\sigma \Delta_4\sigma
\right]\right\}\ ,
\end{equation}
where
\begin{equation}
\label{deltafour}
\Delta_4=\Box^2+2R^{\mu\nu}\nabla_\mu\nabla_\nu
-\frac{2}{3}R\Box+\frac{1}{3}\nabla^\mu R\nabla_\mu\ .
\end{equation}

At this point the reader will wonder whether the two procedures described above
lead to different physical predictions or not.
If the metric and dilaton are treated as classical external fields,
but we allow them to be transformed, the two quantization procedures yield equivalent physics.
In the Weyl--invariant procedure one has the freedom of choosing a gauge
where $\chi=\mu$ and in this gauge all the results reduce to those of the standard procedure.
In particular we observe that the trace of the energy-momentum tensor derived from the two
actions $\GammaI$ and $\GammaII$ are the same.
This follows from the fact that
\be
\int dx\sqrt{g}\,
\frac{\delta\Gamma_{WZ}}{\delta g_{\mu\nu}}
\Bigg|_{(g,\chi=\mu)}2\omega g_{\mu\nu}=0
\ee
which follows from (\ref{wz}).
We will see this in an explicit example in section 4.2.1.

On the other hand if we assume that the metric (and dilaton) are going to be quantized too,
the answer hinges on the choice of {\it their} functional measure.
We defer the discussion of this point to section 7.

\section{The Effective Average Action of free matter fields coupled to an external gravitational field}

In this section we introduce a generalization of the effective action,
called Effective Average Action (EAA), which depends on a scale $k$
having the meaning of infrared cutoff.
The main virtue of this definition is that there exists a simple formula
for the derivative of the EAA with respect to $k$, called
the Functional RG Equation (FRGE) or Wetterich equation \cite{wetterich}.
Its generalization to the context of gravity has been presented in \cite{reuter1}.
The full power of the FRGE, which is an exact equation, becomes manifest when one
considers interacting fields.
In this section we shall familiarize ourselves with the FRGE in the context of
free matter fields coupled to an external metric and dilaton,
where the one--loop approximation is exact.
We leave the discussion of interacting matter to the section 5.

\subsection{The EAA and its flow at one loop}

The definition of the EAA follows the same steps of the definition of the
ordinary effective action, except that one modifies the bare action
by adding to it a cutoff term $\Delta S_k(\phi)$ that is quadratic in the fields 
and therefore modifies the propagator without affecting the interactions. 
Using the notation of \eq{gaussianaction}, the cutoff term is:
\be
\Delta S_k(g_{\mu\nu},\phi)=
\frac{1}{2}\,\cG\!\left(\phi,\frac{R_k(\Delta)}{\mu^2}\phi\right)
=\frac{1}{2}\frac{k^2}{\mu^2}\sum_n a_n^2 r\left(\frac{\lambda_n}{k^2}\right)\ ,
\ee
where we have written the cutoff (which has dimension of mass squared) as
$R_k(z)=k^2 r(z/k^2)$.
The kernel $R_k(z)$ is arbitrary, except for the general requirements of being
a monotonically decreasing function both in $z$ and $k$,
tending fast to zero for $z \gg k^2$ and to $k^2$ for $z\to0$.
These conditions ensure that the contribution to
the functional integral of field modes with momenta $q^2\ll k^2$ are
suppressed, while the contribution of field modes with momenta $q^2\gg k^2$ are unaffected.

We define a $k$-dependent generating functional $W_k$ by
\be
e^{-W_k(g_{\mu\nu},j)}= \int D\phi\,\exp\left(-S(g_{\mu\nu},\phi)
-\Delta S_k(g_{\mu\nu},\phi)-\int dx\, j\phi\right) \ .
\ee
The EAA is obtained by Legendre transforming, and then subtracting the cutoff:
\begin{equation}
\label{eaa}
\Gamma_{k}(g_{\mu\nu},\phi)=W_{k}(g_{\mu\nu},j)-\int dx\, J\phi
-\Delta
S_{k}(g_{\mu\nu},\phi)\,.
\end{equation}
Note that since $R_k\to 0$ when $k\to0$, the EAA becomes the ordinary effective action
in this limit.

The evaluation of the EAA for Gaussian matter fields, conformally coupled to a metric,
follows the same steps that led to (\ref{oneloop}).
The only differences are the replacement of $S$ by $S+\Delta S_k$ and hence of
the ``inverse propagator'' $\Delta$ 
by the ``cutoff inverse propagator'' $P_k(\Delta)=\Delta+R_k(\Delta)$,
and in the end the subtraction of $\Delta S_k$.
The result is
\be
\label{eaaI1loop}
\GammaI_k(g_{\mu\nu},\phi)=
S(g_{\mu\nu},\phi)+\frac{1}{2}\Tr\log\left(\frac{P_k(\Delta)}{\mu^2}\right)\ .
\ee
We used here the superscript I to denote that this EAA has been obtained by 
using the standard measure and reduces to $\GammaI$ for $k=0$.
We would like now to define a Weyl--invariant form of EAA, to be called $\GammaII_k$
in analogy to the effective action $\GammaII$ discussed previously.

The first step is to clarify the meaning of the cutoff $k$ in this context.
In the usual treatment, {\it i.e.} in a non--gravitational context, 
$k$ is a constant with dimension of mass.
In the present context these two properties are contradictory.
A quantity that has a nonzero dimension cannot generally be a constant:
it can only be constant in some special gauge.
This means that the cutoff must be allowed to be a generic non-negative function on spacetime.

Now we must give a meaning to the notion that the couplings depend on the cutoff.
In a Weyl--invariant theory all couplings are dimensionless,
and the only way they can depend on $k$ is via the dimensionless combination $u=k/\chi$.
Note that by definition the dilaton cannot vanish anywhere,
whereas the cutoff should be allowed to go to zero.
So $u$ is a non-negative dimensionless function on spacetime.
This raises the question of the meaning of a running coupling whose argument
is itself a function on spacetime.
In order to avoid such issues we will restrict ourselves to the case
when $u$ is a constant, in other words the cutoff and the dilaton are proportional.

With this point understood, the evaluation of the EAA with the Weyl--invariant measure
is very simple: as in section 3.3 we just have to replace $\mu$ by $\chi$
\bea
\label{eaaII1loop}
\GammaII_u(g_{\mu\nu},\phi)&=&
S(g_{\mu\nu},\phi)+\frac{1}{2}\Tr\log\left(\frac{\Delta+R_k(\Delta)}{\chi^2}\right)
\\
&=& S(g_{\mu\nu},\phi)+\frac{1}{2}\Tr\log\left(\cO+u^2 r(u^2\cO)\right)\ .
\eea
In the second line we have reexpressed the EAA as a function of the Weyl--covariant operator 
$\cO=\chi^{-2}\Delta$,
the Weyl--invariant cutoff parameter $u$ and the dimensionless function $r(z/k^2)=R_k(z)/k^2$.
It is manifest that all dependence on $k$ is via $u$ and that $\GammaII_u$ is Weyl--invariant.

\subsection{Calculating the effective action with the FRGE}

In the preceding section we have defined the EAA, an effective action
that depends continuously on a parameter $k$
and reduces to the ordinary effective action for $k=0$.
It can be shown that the EAA satisfies the following FRGE:
\begin{equation}
\label{ERGE}
k\frac{d\Gamma_k}{dk}=
\frac{1}{2}\mathrm{Tr}\left[
\frac{\delta^{2}(\Gamma_{k}+\Delta S_k)}{\delta\phi\delta\phi}\right]^{-1}k\frac{d}{dk}\frac{\delta^2\Delta S_k}{\delta\phi\delta\phi}\ ,
\end{equation}
which is an exact equation holding for any theory \cite{wetterich}.
\footnote{Note that the structure of (\ref{ERGE}) in field space is the trace of a
contravariant two tensor times a covariant two--tensor
(in de Witt notation, $((\Gamma_k^{(2)}+\Delta S_k^{(2)})^{-1})^{ij}(\partial_t\Delta S_k^{(2)})_{ji}$,
where a superscript $(2)$ denotes second functional derivative and $t=\log k$) and is therefore an invariant expression.
In passing from (\ref{ERGE}) to (\ref{ERGEI}) one uses the field space metric $\cG$
to raise and lower indices and transform the covariant and contravariant tensors into mixed tensors,
each of which can be seen as a function of $\Delta$.
In practice this amounts to canceling all factors of $\sqrt{g}$ and $\mu$.}
We will not give the general proof of this equation but we shall derive 
it in the special case of free matter coupled to an external gravitational field.
Before doing this, however, it is convenient to discuss the use of this
equation as a tool to calculate the effective action.

The r.h.s. of the FRGE (\ref{ERGE}) can be regarded as the
``beta functional'' of the theory, giving the $k$--dependence of all the couplings. 
To see this let us assume that $\Gamma_k$ admits a derivative expansion of the form
\begin{equation}
\label{Gammak}
\Gamma_{k}(\phi,g_{i})=
\sum_{n=0}^\infty\sum_{i}g_{i}^{(n)}(k)\calo_{i}^{(n)}\left(\phi\right)\ ,
\end{equation}
where $g_{i}^{(n)}(k)$ are coupling constants and $\calo_{i}^{(n)}$
are all possible operators constructed with the field $\phi$ and
$n$ derivatives, which are compatible with the
symmetries of the theory. 
We have
\begin{equation}
\label{dtGamma}
k\frac{d\Gamma_k}{dk}=\sum_{n=0}^\infty\sum_{i}\beta_{i}^{(n)}\calo_{i}^{(n)}\ ,
\end{equation}
where
$\beta_{i}^{(n)}(g_{j},k)=k\frac{dg_{i}^{(n)}}{dk}=\frac{dg_{i}^{(n)}}{dt}$
are the beta functions of the couplings.
Here we have introduced $t=\log(k/k_{0})$, $k_{0}$ being an arbitrary initial value.
If we expand the trace on the r.h.s. of (\ref{ERGE}) in operators
$\calo_i^{(n)}$ and compare with (\ref{dtGamma}), we can
read off the beta functions of the individual couplings.

The most remarkable property of the FRGE is that 
the trace on the r.h.s. is free of UV and IR divergences.
This is because the derivative of the cutoff kernel goes rapidly to zero for $q^2>k^2$,
and $k$ also acts effectively as a mass.
So, even though the EAA defined above is as ill--defined as the usual effective action,
its $t$--derivative is well--defined.
Given a ``theory space'' which consists of a class of functionals
of the fields, one can define on it a flow without having to worry
about UV regularizations. All the beta functions are finite.
This can be done for any theory, whether renormalizable or not.

Then, one can pick an initial point in theory space, which can be identified 
with the bare action at some UV scale $\Lambda$, and study the trajectory passing
through it in either direction.
The EAA can be obtained by solving the first order differential equation (\ref{ERGE})
and taking the limit $k\to0$.

The issue of the divergences presents itself, in this formulation, when one
tries to move $\Lambda$ to higher energies, which is equivalent to solving
the RG equation for growing $k$.
If the trajectory is renormalizable, all dimensionless couplings remain finite in the limit $k\to\infty$. 
This implies that only the relevant dimensionful coupling diverge, and one expects only a finite
number of these.
The ambiguities that correspond to these divergences are fixed by the choice of RG trajectory,
because the IR limit (i.e. the renormalized couplings) is kept fixed.
On the other hand if some dimensionless coupling diverges (e.g. at a Landau pole)
the theory ceases to make sense there and the trajectory describes an effective low energy
field theory.

Let us now return to the case of free matter in an external gravitational field.
In the preceding section we defined two variants of the EAA:
the ``standard'' EAA $\GammaI_k$ and the Weyl--invariant EAA $\GammaII_u$,
both of which can be written as trace of the logarithm of some
function of the kinetic operator.
These expressions are formal, because they contain divergences and 
need to be regularized.
In the case of $\GammaII_u$ this can be done in a Weyl--invariant way
by using an UV cutoff that is a multiple of the dilaton,
similar to the way we introduced the infrared cutoff $k$.
We do not pursue this here.
Instead, we take the derivative of (\ref{eaaI1loop}) with respect to $k$ and 
using the definition $R_k(\Delta)=k^2r(\Delta/k^2)$ obtain
\begin{align}
\label{ERGEI}
k\frac{d\GammaI_k}{dk}=&
\,\frac{1}{2}\mathrm{Tr}\left(\frac{1}{\Delta+R_{k}(\Delta)}k\frac{dR_k(\Delta)}{dk}\right)
\nonumber
\\
=&\,\mathrm{Tr}\frac{r(\Delta/k^2)-(\Delta/k^2)r'(\Delta/k^2)}{(\Delta/k^2)+r(\Delta/k^2)}\ .
\end{align}
It is easy to see, especially using the form in the first line,
that this is a special case of the FRGE \eq{ERGE},
and the fall--off properties of the function $r$ guarantee that the trace on the r.h.s. is finite.

One can repeat this argument in the case of the Weyl--invariant EAA
with little changes, and the flow equation reads
\begin{equation}
\label{ERGEII}
u\frac{d\GammaII_u}{du}=
\mathrm{Tr}\frac{r(\cO/u^2)-(\cO/u^2) r'(\cO/u^2)}{(\cO/u^2)+r(\cO/u^2)}\ .
\end{equation}
In this form the r.h.s. of the FRGE is manifestly Weyl--invariant,
since $u$ is Weyl--invariant and one has the trace of a function of 
a Weyl--covariant operator.
\footnote{Note that $\Delta/k^2=\cO/u^2$ so the r.h.s. of \eq{ERGEI} and \eq{ERGEII} are identical.
The reason for the lack of invariance of the EAA $\GammaI$ (and its derivative)
is the measure which contains the absolute mass scale $\mu$.
If one allowed $\mu$ to be transformed, in the same way as we allow the cutoff $k$
to be transformed, the two actions would be seen to be the same.}

The EAA's $\GammaI_k$ and $\GammaII_u$ are not well--defined functionals,
but their derivatives are well--defined.
As explained above, one can integrate the FRGE
and obtain, in the IR limit, the ordinary effective action.
If one starts from a given Weyl--invariant classical matter action at scale $\Lambda$
and integrates the flow of $k\frac{d\GammaI_k}{d k}$, 
respectively $u\frac{d\GammaII_u}{du}$,
down to $u=0$ one obtains exactly the effective action $\GammaI$, respectively $\GammaII$.
Furthermore, at each $u$, $\GammaII_u$ is obtained from $\GammaI_k$ by the St\"uckelberg trick.
It is instructive to explicitly illustrate these statements in the case of $d=2$ and, 
for the $c$--anomaly, also in the case $d=4$.

\subsubsection{$d=2$: the Polyakov action}
In this section we consider the effective action of a single scalar field \cite{polyakov}.
It has been derived by integrating the FRGE in \cite{codello1}.
The main tool in this derivation is the non--local expansion of the heat kernel in powers of curvature \cite{russian1,coza}.
Keeping terms up to two curvatures one has
\be
\Tr e^{-s\Delta}=\frac{1}{4\pi s}\int d^2x\sqrt{g}\,\left[1+s\frac{R}{6}
+s^2R f_R(s\Delta)R+\ldots\right]\ ,
\ee
where
$$
f_R(x)=\frac{1}{32}f(x)+\frac{1}{8x}f(x)-\frac{1}{16x}+\frac{3}{8x^2}f(x)-\frac{3}{8x^2}\ ;
\qquad
f(x)=\int_0^1 d\xi e^{-x\xi(1-\xi)}\ .
$$
The r.h.s. of (\ref{ERGEI}) can be written, after some manipulations,
$$
k\frac{d\GammaI_k}{dk}=\int ds\, \tilde h(s)\Tr\,e^{-s\Delta}\ ;
\qquad
h(z)=\int_0^\infty ds\,\tilde h(s) e^{-s\,z}\ ,
$$
where $\tilde h(s)$ is the Laplace anti-transform of $h(z)=\frac{\partial_tR_k(z)}{z+R_k(z)}$.
Using the explicit cutoff $R_k(z)=(k^2-z)\theta(k^2-z)$, we have simply $h(z)=2k^2\theta(k^2-z)$
and the integrals give
\bea
k\frac{d\GammaI_k}{dk}&=&
\int d^{2}x\sqrt{g}\Biggl[\frac{k^2}{4\pi}
+\frac{1}{24\pi}R
\\
&&+\frac{1}{64\pi}R\frac{1}{\Delta}
\left(\sqrt{\frac{\tDelta}{\tDelta-4}}
-\frac{\tDelta+4}{\tDelta}\sqrt{\frac{\tDelta-4}{\tDelta}}\right)\theta(\tDelta-4)R
\Biggr]+O(R^3)
\nonumber
\eea
with $\tDelta=\Delta/k^2$.
On the other hand, keeping terms at most quadratic in curvature, the EAA can be written in the form
\be
\label{2dansatz}
\GammaI_k=\int d^{2}x\sqrt{g}\left[a_{k}+b_{k}R+R\,c_{k}(\Delta)R\right]+O\left(R^{3}\right)
\ee
where $c_{k}\left(\Delta\right)$ is a nonlocal form-factor which,
for dimensional reasons, can be written in the form $c_{k}\left(\Delta\right)=\frac{1}{\Delta}c(\tDelta)$.
The beta functions of $a_k$, $b_k$ and $c_k$ are then
\be
\partial_t a_k=\frac{k^2}{4\pi}\ ;\quad
\partial_t b_k=\frac{1}{24\pi}\ ;\quad
\partial_t c=\frac{1}{64\pi}
\left(\sqrt{\frac{\tDelta}{\tDelta-4}}
-\frac{\tDelta+4}{\tDelta}\sqrt{\frac{\tDelta-4}{\tDelta}}\right)\theta(\tDelta-4)
\ee
In order to obtain the effective action,
one integrates this flow from some UV scale $\Lambda$, that can later be sent to infinity,
down to $k=0$.
Setting $a_\Lambda=\frac{\Lambda^2}{4\pi}$, one has $a_k=\frac{k^2}{4\pi}$ and therefore
the renormalized cosmological term vanishes in the IR limit.
The Hilbert term has a logarithmically running coefficient
$b_k=b_\Lambda-\frac{1}{24\pi}\log\frac{\Lambda}{k}$.
We will not consider this term in the following because it is topological.
We assume that $c_k$ vanishes at $k\rightarrow\infty$, since the UV action only contains the matter terms.
The integral over $k$ is finite even in the limit $\Lambda\to\infty$, and one finds
\be
\label{ck}
c(\tDelta)=-\frac{1}{96\pi}
\frac{\sqrt{\widetilde\Delta-4}(\widetilde\Delta+2)}{\widetilde\Delta^{3/2}}
\theta(\widetilde\Delta-4)\ .
\ee

The explicit form of $c_k$ can be found also employing the mass cutoff $R_{k}(z)=k^2$, in which case
the computation can also be done analytically, giving
\be
c(\tilde\Delta)=-\frac{1}{16\pi}\left[
\frac{1}{6}-\frac{1}{\tilde\Delta}
+\frac{\mathrm{Arctanh}\left(\sqrt{\frac{\tilde\Delta}{\tilde\Delta+4}}\right)}
{{\tilde\Delta}^{3/2}\sqrt{\tilde\Delta+4}}\right]
\ee
and with the exponential cutoff $R_k(z)=\frac{z}{exp(\frac{z}{k^2})-1}$, 
in which case it is computed numerically.
All three give the same qualitative running, as depicted in figure 2.
In the limit $k\to 0$ one obtains, in all cases, the Polyakov action:
\footnote{Using this action in (\ref{wz}) one recovers the WZ action (\ref{wz2}).
Conversely, the Polyakov action can be obtained from the WZ action by
using the equation of motion for $\sigma$.}
\be
\label{polyakov}
\GammaI(g_{\mu\nu})=-\frac{1}{96\pi}\int d^{2}x\sqrt{g}R\frac{1}{\Delta}R\ .
\ee

\begin{figure}
[t]\center
{\resizebox{0.6 \columnwidth}{!}
{\includegraphics{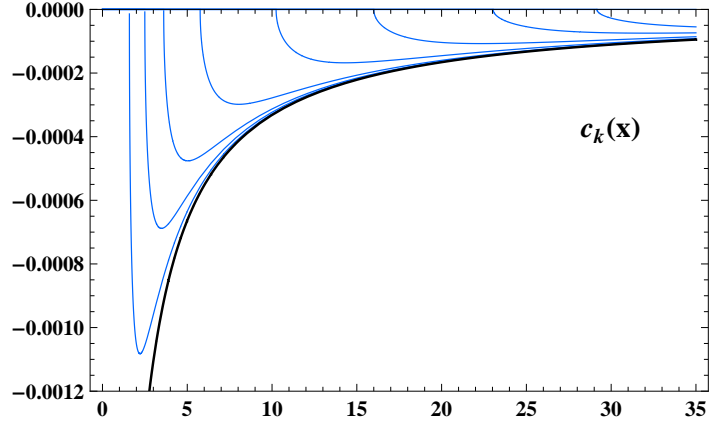}}
\caption{\label{fig:flowmatter}Shape of the form factor $c_k(x)$ of eq. \eq{ck} as a function of $x=\Delta$ for different values of $k$. The thick line shows the case $k=0$ (Polyakov action).
}}
\end{figure}

The function $c_k$ admits a series expansion
$c_k(\Delta)=\frac{1}{k^2}\sum_{n=1}^\infty c_n\frac{k^{2n}}{\Delta^n}$.
Then, one can explicitly perform the variation with respect to the metric
and obtain the energy--momentum tensor. In particular,
conformal variation of $\GammaI_k$ gives the $k$--dependent trace anomaly:
\be
\label{silvia}
\langle T^\mu_\mu\rangle_k^I =
-\frac{2}{\sqrt g}g_{\mu\nu}\frac{\delta\GammaI_u}{\delta g_{\mu\nu}}
= -4 c(\tilde{\Delta})R
-\frac{2}{k^2}\sum_{n=0}^\infty\sum_{k=1}^{n-1}c_n
\left(\frac{1}{\tilde\Delta^k}R\right)
\left(\frac{1}{\tilde\Delta^{n-k}}R\right)\ .
\ee
We observe that the integrated trace anomaly (which is related to the variation of the EAA
under a global scale transformation) can be written more explicitly
\be
\int dx\sqrt g\,\langle T^\mu_\mu\rangle_k = 
\int dx\sqrt g\,\left(-4 c(\tilde{\Delta})R
+\frac{2}{k^2} R\, c'(\tilde{\Delta})R\right)\ .
\ee

For a fixed momentum $\Delta$ the linear term of the trace anomaly grows monotonically as $k$ decreases,
from zero at infinity to its canonical value at $k=0$. The second term shows a nontrivial flow for $k\neq 0$, going to zero both in the UV and IR.

\begin{figure}
[t]\center
{\resizebox{0.6 \columnwidth}{!}
{\includegraphics{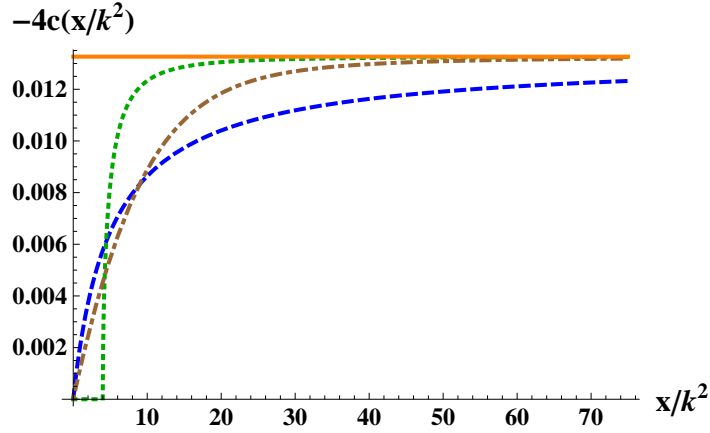}}
\caption{\label{fig:flowmatter}Flow of the trace anomaly as a function of $x/k$ for 
fixed $x=\Delta$, as given by the function $-4c(\frac{x}{k^2})$ of equation \eq{silvia}.
Note that the origin is the UV limit.
The three curves correspond to optimized cutoff (green, dotted), exponential cutoff (brown, dot-dashed) and mass-type cutoff (blue, long -dashed); also plotted is the $k=0$ asymptotic value of the trace anomaly (orange).
}}
\end{figure}

Let us now come to the effective action $\GammaII$.
Using the Weyl--invariant measure, the effective action is given by the 
determinant of the dimensionless
operator $\cO=\hat\Delta=\frac{1}{\chi^2}\Delta$, which can be identified with $\Delta_{\hat g}$,
the operator constructed with the dimensionless, Weyl--invariant metric $\hat g_{\mu\nu}=\chi^2 g_{\mu\nu}$.
Therefore, as already discussed, $\GammaII$ differs from $\GammaI$ just in the replacement of 
$\mu^2 g_{\mu\nu}$ by $\chi^2 g_{\mu\nu}$.
We have to generalize this for finite $k\not=0$. 
As discussed above, we assume that the cutoff is a constant multiple of the dilaton: $k=u\chi$.
Neglecting the $a$-- and $b$--terms,
the effective average action can then be written in the manifestly Weyl--invariant form
\be
\GammaII_{u}(g_{\mu\nu},\chi)=\int d^2x\sqrt{g}\,
\cR\, \frac{1}{\chi^2\cO}c\left(\frac{\cO}{u^2}\right) \cR\ ,
\ee
with the same function $c$ given in (\ref{ck}).
In particular the Weyl--invariant version of the
Polyakov action is obtained in the limit $u\to0$:
\be
\GammaII(g_{\mu\nu},\chi)=
-\frac{1}{96\pi}\int d^2x\sqrt{g}\,
\cR\, \frac{1}{\chi^2\cO} \cR\ .
\ee

We have claimed in the end of section 3.4 that the trace of the energy--momentum tensors
computed from $\GammaII$ and $\GammaI$ coincide in the gauge $\chi=\mu$.
This statement actually holds also for $k\not=0$.
A direct calculation yields
\be
\label{bruno}
\langle T^\mu_\mu\rangle_u^\mathrm{II}=
-\frac{2}{\sqrt g}g_{\mu\nu}\frac{\delta\GammaII_u}{\delta g_{\mu\nu}} 
= -4c\left(\frac{\cO}{u^2}\right)\cR
-\frac{2}{u^2\chi^2}\sum_{n=1}^\infty\sum_{k=1}^{n-1}c_n
\left(\frac{u^{2k}}{\cO^k}\cR\right)
\left(\frac{u^{2(n-k)}}{\cO^{n-k}}\cR\right)
\ee
One can verify that this is also equal to $\frac{1}{\sqrt g}\chi\frac{\delta\GammaII_u}{\delta\chi}$,
thereby obtaining an explicit check of the general statement \eq{olga}.
It is also interesting to observe that if we think of $\GammaII_u$
as a function of $k$, $\chi$ and $g_{\mu\nu}$,
and vary each keeping the other two fixed, the metric variation is
again given by equation \eq{bruno}, the $\chi$ variation
gives the first term in the r.h.s. of \eq{bruno}
and the $k$ variation gives the second term.
We also note that the ``beta functional'' can be written in general as
\be
u\frac{d\GammaII_u}{du}=-\int dx\sqrt g\,\frac{2}{u^2\chi^2}
\cR\, c'\left(\frac{\cO}{u^2}\right)\cR\ .
\ee

\subsubsection{$d=4$: the $c$--anomaly action}

One would like to repeat the analysis of the previous section in $d=4$, 
to the extent that this is possible.
The main difference is that while in $d=2$ the Polyakov action is the full effective action,
in $d=4$ there are terms with higher powers of curvature.
The analysis then has to be limited to the first few terms of the expansion in curvatures:
\be
\GammaI_k=\int d^4x\sqrt{g}\left[a+ b R +Rf_R(\Delta)R
+C_{\mu\nu\rho\sigma}f_C(\Delta)C^{\mu\nu\rho\sigma}+O(R^3)\right]
\ee
where $a$, $b$, $f_R$ and $f_C$ depend on $k$. Such running structure functions have been already computed in \cite{codello2}. 
It was found that the form factor $f_R (\Delta)$ tends to zero in the IR limit,
whereas $f_C (\Delta)$ approaches the standard one-loop EA at two curvatures for $k\to0$:
\begin{equation}
\GammaI = -\frac{1}{2}\frac{1}{(4\pi)^2} \int d^4 x \sqrt{g}\,
\frac{N_0+6 N_{1/2}+12 N_1}{120}\, 
C_{\mu \nu \rho \sigma} \log \left(\frac{\Delta}{k_0^2} \right) C^{\mu \nu \rho \sigma} 
+\ldots
\label{EA d=4 k=0}
\end{equation}
where $\Delta= -\nabla^2$ and $k_0$ is an arbitrary scale.
Here we are neglecting the local terms, whose coefficients are arbitrary and can be tuned to zero.
To connect this result with standard one loop EAs computed in \cite{russian1} it is sufficient to change the basis expansion from powers of $(R,C_{\mu\nu\rho\sigma})$ and their derivatives to 
powers of $(R,R_{\mu\nu})$ and their derivatives.
(This requires expressing the Riemann tensor as an infinite nonlocal series in the Ricci tensor.)

An EA of the form (\ref{EA d=4 k=0}) had been suggested by Deser and Schwimmer \cite{deser} as the source of the $c$--anomaly, namely the terms proportional to $C^2_{\mu \nu \rho \sigma}$ in \eq{traceanomaly4}. 
This action (in contrast to the Riegert action discussed below) 
also produces the correct flat spacetime limit for the correlation 
functions of the energy momentum tensor 
$\langle T_{\mu \nu} T_{\rho \sigma} \rangle$ \cite{erdmenger}.

In the basis of the tensors $(R,R_{\mu\nu})$ the terms cubic in curvature are known explicitly \cite{russian2}.
When the Riemann squared term in the anomaly is expanded in an infinite series in
$(R,R_{\mu\nu})$, the action of \cite{russian2} correctly reproduces the first
terms of this expansion \cite{russian3}.
In order to reproduce the full anomaly (both $c$-- and $a$--terms)
one would need also terms in the effective action of order higher than three.

It is possible to write closed form actions that generate the full anomaly.
A functional that generates the $c$-anomaly has been given already in \eq{EA d=4 k=0}.
Another action that gives both $c$- and $a$--anomaly is the Riegert action \cite{riegert}
\begin{equation}
\label{riegert}
W(g_{\mu\nu})=\int dx\,\sqrt{g} 
\frac{1}{8} \left(E-\frac{2}{3}\Box R\right)
\Delta_4^{-1}
\left[2c\,C^2-a\left(E-\frac{2}{3}\Box R\right)\right]+\frac{a}{18} R^2\, .
\end{equation}
It has the drawback that it gives zero for the flat spacetime limit
of the correlator of two energy--momentum tensors.
This does not mean, however, that one cannot write the full effective action
as the sum of the Riegert action and Weyl--invariant terms,
because one can write the Deser--Schwimmer action as the Riegert action (with $a=0$)
plus Weyl--invariant terms.
In this case the energy--momentum correlator would come from the Weyl--invariant terms,
as we shall see below.

The relation between the Wess--Zumino term \eq{wz4} and the Riegert
action \eq{riegert} is very similar to the one between the two--dimensional
Wess--Zumino action \eq{wz2} and the Polyakov action \eq{polyakov}:
using the Riegert action in (\ref{wz}) one recovers the WZ action (\ref{wz4}).
Unlike the two--dimensional case, however, the converse procedure is not unique.
The general  idea is to replace the dilaton $\chi=\mu e^\sigma$,
which in the WZ action is treated as an independent variable, 
by a functional of the metric $g_{\mu\nu}$ having the right transformation properties.
One choice, which has been proposed in \cite{fv,russian4} is 
\begin{equation}
\label{marta}
\sigma (g_{\mu\nu})=  \log \left(1-\frac{1}{\Delta +R/6} \frac{R}{6}\right) .
\end{equation}
Another possibility is
\begin{equation}
\label{sandra}
\sigma (g_{\mu\nu})=-\frac{1}{4}\frac{1}{\Delta_4}\left(E+\frac{2}{3}\Delta R+b\,C^2\right) ,
\end{equation}
where $b$ is an arbitrary constant. In both cases 
$\sigma(g_{\mu\nu})\mapsto\sigma(g_{\mu\nu})-\log\Omega$
under a Weyl transformation.
Note that \eq{sandra}, for $b=c$ is the equation of motion for the dilaton
coming from the WZ action \eq{wz4}, while for $b=0$ it is the equation of motion coming from
the $a$--term of the WZ action.
The latter choice exactly reproduces \eq{riegert}; other choices of $b$ give the Riegert action
plus Weyl--invariant terms, while \eq{marta} gives another form of the anomaly functional.

One can obtain some additional information on the effective action $\GammaI$
by using these formulae in
\begin{equation}
\GammaI(g_{\mu\nu})=\GammaII(g_{\mu\nu},\chi)-\Gamma_{WZ}(g_{\mu\nu},\chi) \ ,
\label{rel GI GII}
\end{equation}
where the first term in the r.h.s. is Weyl--invariant by construction
and the anomaly comes entirely from the second term.
For example if we use \eq{sandra} with $b=0$, the second term exactly reproduces
the Riegert action and the correlator of two energy--momentum tensors must come from
the first term.
We know already that it must contain the term
\begin{equation}
\label{claudia}
\GammaII(g_{\mu\nu},\mu e^{\sigma(g_{\mu\nu})}) = -\frac{1}{2}\frac{1}{(4\pi)^2} \int d^4 x \sqrt{g}\,
\frac{N_0+6 N_{1/2}+12 N_1}{120}\, 
\cC_{\mu \nu \rho \sigma} \log \left(\frac{\cO}{u_0^2} \right) \cC^{\mu \nu \rho \sigma} +\ldots
\end{equation}
where $\cC_{\mu\nu\rho\sigma}$ is the Weyl tensor constructed with the metric $e^{2\sigma(g)}g_{\mu\nu}$.
Expanding this to second order in the curvature of $g_{\mu\nu}$ one reobtains as a leading term
the action \eq{EA d=4 k=0}.
The lack of Weyl--invariance of that action is compensated by higher terms in the expansion.
This shows that there is no contradicton between the presence of the Riegert and the Deser--Schwimmer terms in the effective action $\GammaI$, and the flat space limit of
energy--momentum tensor correlators.
Thus there is also no disagreement with \cite{mottola1} and with \cite{russian4}.

\begin{figure}
[t]\center
{\resizebox{0.6 \columnwidth}{!}
{\includegraphics{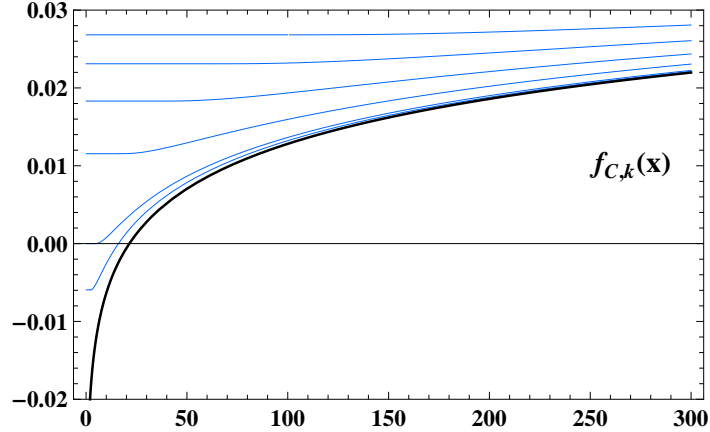}}
\caption{\label{fig:flowmatter}Shape of the form factor $f_{\cC}$ of eq. \eq{indeaa4}
in the case of a single scalar field, as a function of $x = \Delta$ for different values of $k$. The thick line shows the case $k=0$.
}}
\end{figure}

Finally, we can write the explicit form of the interpolating EAA.
For a scalar field we have \cite{codello2}
\begin{eqnarray}
\label{indeaa4}
\GammaI_k(g_{\mu\nu})&=& 
-\frac{1}{2 (4 \pi)^2} \int d^4x\sqrt{g}\, 
C_{\mu\nu\rho\sigma}\Biggl\lbrace  
\frac{1}{120}\log \left(\frac{k^2}{k_0^2} \right)
\nonumber\\
&+& \theta \left( \tilde{\Delta} -4 \right)
\Biggl[-\frac{1}{120}\log \left(\frac{k^2}{k_0^2} \right)
-\frac{4  \sqrt{\tilde{\Delta}-4} \sqrt{\tilde{\Delta}}}{75 \tilde{\Delta}^3}
+ \frac{11 \sqrt{\tilde{\Delta}-4} \sqrt{\tilde{\Delta}}}{225 \tilde{\Delta}^2}
\nonumber \\
&-&\frac{23}{900} \sqrt{1-\frac{4}{\tilde{\Delta}}}
+\frac{1}{120} \log \left(\frac{\Delta}{2 k^2_0}
\left(\sqrt{1-\frac{4}{\tilde{\Delta}}}+1\right)-\frac{k^2}{k_0^2}\right)\Biggr]\Biggr\rbrace
C^{\mu\nu\rho\sigma}+\ldots\ ,
\end{eqnarray}
Notice that the first logarithm in the bracket is both UV and IR divergent,
and also note that the Heaviside theta is zero when $k$ is sufficiently large.
Thus, the UV divergence is present and must be removed by renormalization,
whereas the IR divergence is automatically canceled by the second logarithm.
The form factor $f_C(x)$, for fixed $x$, is plotted in figure 3.
Similar formulas, but with different coefficients, hold also for fermions and gauge fields.
In the limit $k\to0$ they all reduce to \eq{EA d=4 k=0}

The calculation of $\GammaII$ follows the same lines. 
There are two running structure functions $f_\cR$ and $f_\cC$.
The explict form of $f_{\cC}$ for a scalar field is 
\begin{eqnarray}
\GammaII_{u}(g_{\mu\nu},\chi)&=& 
-\frac{1}{2}\frac{1}{(4\pi)^2} \int d^4x\sqrt{g}\,\cC_{\mu\nu\rho\sigma}\Biggl\lbrace  
\frac{1}{120}\log u^2
\nonumber
\\
&+& \theta \left( \frac{\cO}{u^2}-4 \right)
\Biggl[-\frac{1}{120}\log u^2
-\frac{4 u^6 \sqrt{\frac{\cO }{u^2}-4} \sqrt{\frac{\cO }{u^2}}}{75 \cO ^3}
+ \frac{11 u^4
\sqrt{\frac{\cO }{u^2}-4} \sqrt{\frac{\cO }{u^2}}}{225 \cO ^2}
\nonumber
\\
&-&\frac{23}{900} \sqrt{1-\frac{4 u^2}{\cO }}
+\frac{1}{120} \log \left(\frac{\cO}{2}
\left(\sqrt{1-\frac{4 u^2}{\cO}}+1\right)-u^2\right)\Biggr]\Biggr\rbrace
\,\cC^{\mu\nu\rho\sigma}+\ldots
\end{eqnarray}
The same computation can be repeated in the case of fermions and vectors and a different interpolating function can be found. When $u \rightarrow 0$ we get back equation \eq{claudia}.

\section{Interacting matter fields}

In the preceding sections we have shown that there exists a quantization procedure
such that the effective action which is obtained by integrating out free (Gaussian) matter fields
remains Weyl invariant.
The proof was simple because the integration over matter was Gaussian.
Here we generalize the result to the case when there are matter interactions.

As in the preceding section, we begin by considering the case when the
initial matter action is Weyl invariant even without invoking a coupling to the dilaton.
This is the case for massless, renormalizable quantum field theories such
as $\phi^4$, Yang-Mills theory and fermions with Yukawa couplings.
The interactions are of the form $S_{int}(g_{\mu\nu},\Psi_a)=\lambda\int d^4x\sqrt{g}\,\calL_{int}$
where $\calL_{int}$ is a dimension $d$ operator and $\lambda$ is dimensionless.
Interactions generate new anomalous terms over and above those that we have
already considered for Gaussian matter.
The trace anomaly of free matter vanishes in the limit of flat space,
but this is not true for interacting fields: the trace is then proportional to the beta function.
For the interaction term given above one has in flat space
\be
\label{enzo}
\int dx\,\omega \langle T^\mu{}_\mu\rangle=
-\delta_\omega S_{int}=
\int d^4x\,\omega\,\beta_\lambda\,\calL_{int}
\ee
where $\beta_\lambda=k\frac{d\lambda}{d k}$.
(This is somewhat similar to equation \eq{tb}, but there is a sign difference
due to the fact that $\mu$ does not play the role of a sliding scale in section 3.)

We want to study the effective action of this theory,
which is obtained by integrating out the matter fields.
In order to be able to make non--perturbative statements we will use 
the FRGE as a machine for calculating the effective action,
as discussed in the introduction and exemplified by the calculations in sections 4.2.1 and 4.2.2.
The general idea is to begin with some Weyl--invariant bare action at some scale and to integrate the RG flow. 
If the ``beta functional'' is itself Weyl--invariant,
the action at each scale will be Weyl--invariant.
The effective action, which is obtained by letting $k\to0$, will also be Weyl--invariant.

This statement is seemingly in contrast with \eq{enzo}, which implies that Weyl invariance
can only be achieved when all beta functions are zero.
How can one maintain Weyl--invariance along a flow?
The trick is to consider the flow as dependence of $\lambda$ on the
dimensionless parameter $u=k/\chi$.
We assume $u$ to be constant to avoid issues
related to the interpretation of a coupling depending on a function.
Since $u$ is Weyl--invariant, also $\lambda(u)$ is.
This is very much in the spirit of Weyl's geometry, where the dilaton
is interpreted physically as the unit of mass and $u$ is the cutoff
measured in the chosen units.

We now see that with this definition of RG, the running of couplings does not in itself
break Weyl invariance.
In the spirit of Weyl's theory the dilaton is taken as a reference scale
and the couplings are functions of $u$.
Since $u$ is Weyl--invariant,
\be
\delta_\omega S_{int}=0\ ,
\ee
even when the beta function $\beta_\lambda=u\frac{d\lambda}{d u}$ is not zero.
It is important to stress that this should not be interpreted as vanishing 
trace of the energy--momentum tensor.
We argued in section 3.4 that the energy--momentum tensor is the same whether
one uses the standard or the Weyl--invariant measure.
That argument is not restricted to non--interacting matter and applies here too.
So, as in the case of free matter fields discussed at the end of the preceding section,
the physical content of the Weyl--invariant theory is exactly the same as in the usual formalism.
The recovery of Weyl--invariance is due to additional terms that involve the
variation of the action with respect to the dilaton.

Let us return now to the issue of the Weyl--invariance of the flow.
The r.h.s. of the FRGE is given in (\ref{ERGE})
as a trace of a function involving the Hessian $\calH$.
As was the case with the determinants of the preceding section,
such a trace can only be defined by using a metric in field space.
It is therefore more appropriate, and more convenient,
to regard the argument of the trace as a function of the differential
operator $\calO$, related to the Hessian as in (\ref{raise}).
Now, the part of the action that is quadratic in the fields coincide
with the free actions considered in the previous section and is Weyl invariant.
It gives the Weyl-covariant operators $\calO$ of the preceding section.
The new interaction terms are also Weyl invariant, and they will
modify the operator $\calO$ by terms that are also Weyl-covariant.
Therefore, the r.h.s. of the FRGE is again the trace of some function of
a Weyl-covariant operator, and therefore is again Weyl invariant.
Since the beta functional is Weyl--invariant, if we start from some
initial condition that is Weyl invariant we will remain within the
subspace of theories that are Weyl--invariant.
The effective action, which is obtained as the limit of the flow
for $k\to 0$, will also be Weyl invariant.

The advantage of the calculation based on the FRGE is that it extends
easily to arbitrary theories.
Let us begin by considering the addition of masses, which break
Weyl invariance at the classical level but remains within the scope
of renormalizable theories.
Applying the St\"uckelberg trick we can convert all mass terms to interactions 
with the background dilaton, thus reinstating Weyl invariance at the classical level.
For example, in the case of a massive scalar field $\phi$ the mass term is written as
$g\chi^2\phi^2$, for some dimensionless coupling $g$.
This becomes a genuine mass term in the gauge where $\chi=\mu$.
Then we can repeat the preceding argument.
The only difference is that now the dilaton will be present in the action
from the beginning, whereas if one had started from a Weyl--invariant theory, 
the coupling to the dilaton
would only arise in the course of the flow as a consequence of its presence in $\cO$.

Finally, we can relax all constraints on the functional form of the action $S(g_{\mu\nu},\psi_a,g_i)$.
Using the St\"uckelberg trick, as discussed in the end of section 3,
we can construct an action $\hat S(g_{\mu\nu},\chi,\psi_a,\hat g_i)$.
Because $\hat S$ is Weyl--invariant, the resulting Hessian has 
well-defined, homogeneous transformation properties
and the operator $\calO$, defined in section 3.3 by using a Weyl--invariant metric $\calG$, 
is Weyl--covariant.
The r.h.s. of the FRGE is a trace of a function of this operator and therefore is Weyl--invariant.

Let us suppose that we know the form of the action $S$ at some (constant) cutoff $k=u\mu$.
It gives rise, via its flow, to an effective action $\Gamma$.
Construct the Weyl--invariant starting action $\hat S$ and take it as
initial point of the flow at cutoff $k=u\chi$ (which could now be some function of position).
Flowing towards the IR from this starting point is guaranteed to lead
to an effective action $\hat\Gamma$ that is still Weyl invariant.
When $\hat\Gamma$ is evaluated at constant $\chi=\mu$ it agrees with the
effective action $\Gamma$ evaluated in the Weyl--non--invariant flow.
We thus see that {\it quantization commutes with the St\"uckelberg trick}.

As mentioned earlier, renormalizability is not required for these arguments, 
because the FRGE is UV finite.
Divergences manifest themselves when one tries to solve for the flow towards large $k$.
The question whether this theory has a sensible UV limit can be answered
by studying the flow for increasing $u$.
If the trajectory tends to an UV fixed point it is called a ``renormalizable''
or ``asymptotically safe'' trajectory.
If instead the trajectory diverges in the UV, it describes an effective field theory
with an UV cutoff scale.
We will address in section 7 the meaning of a fixed point in a theory space
consisting entirely of Weyl--invariant actions.


\section{Dynamical gravity}


Until now we have considered matter fields coupled to an external gravitational
field, which is described either by a metric or by a metric and a dilaton.
\footnote{For the sake of coupling to spinor fields one should use a frame field
rather than a metric. This complication is not relevant for our purposes
and will be ignored. We refer to \cite{hr,dp} for a recent discussion.}
We would like to extend our results also to the case when gravity is dynamical.
This means that we have to be able to ``quantize gravity''.
Contrary to what is often stated, there exists a perfectly well defined
and workable framework that allows us to compute quantum gravitational effects:
it is the framework of effective field theories.
By using the background field method, general relativity reduces to
a (perturbatively nonrenormalizable) theory of a spin two field
propagating on a curved manifold, not unlike the general theories
discussed in the previous section.
\footnote{Ordinary perturbation theory is a special case where the background is flat.}
As discussed for example in \cite{background}, 
the background field method actually guarantees that
the theory is ``background independent'' in the sense that no background
plays a special role.
This scheme has the limitation that in trying to calculate the effective action
one encounters infinitely many divergences, each requiring a physical measurement
to fix the value of the corresponding counterterm.
This means that the theory can be adjusted to fit essentially any experimental
result and is not predictive.
In practice this is not as bad as it seems. As long as one restricts oneself to
experiments at energies below the UV cutoff of the theory,
a finite number of loops is sufficient to describe the data with a predefined precision.
Thus, only finitely many divergences are encountered and one could
test the theory by comparing it to a number of experiments that is greater
than the number of divergences.
This logic has been quite successful in our understanding of strong interactions
at low energy.

If this is still regarded as too unsatisfactory, one can entertain the possibility
that the theory is on a renormalizable trajectory and therefore can be continued to indefinitely large energies.
The advantage of such a situation is that if the attraction basin of the
fixed point is finite dimensional, it places infinitely many
constraints on experiments at any energy scale, and is therefore
highly predictive.
\footnote{This is the logic that led to the standard model of particle physics.}
This possibility, however, is not essential for our main result.
The main fact is that standard quantum field theoretic methods can be used
to describe a quantum field theory of gravity
which is at least an effective field theory with a limited energy domain
of applicability and in the most optimistic scenario may hold up to indefinitely high energy
and have a finite number of free parameters.

\subsection{Weyl--covariant formalism for quantum gravity}

With these motivations we now consider a generic theory of gravity
based on an action $S$ which is a diffeomorphism- and Weyl--invariant
functional of a metric $g_{\mu\nu}$ and a dilaton $\chi$.
By the discussion in section 3, there is a one-to-one correspondence
between such functionals and diffeomorphism-invariant functionals of a metric alone.
If there were some matter fields that have already been integrated out,
the corresponding effective action need not be considered separately
and is included here in the gravitational action.

The metric and dilaton now have to be expanded as the sum of a background and a quantum part.
Since in the following we will have to refer to the backgrounds much more often than
to the full background plus quantum fields, for typographical simplicity
we choose to call $\bar g$ and $\bar\chi$ the full quantum fields,
$g$ and $\chi$ the background fields, $h$ and $\eta$ the quantum fields. Thus
\be
\bar g_{\mu\nu}=g_{\mu\nu}+h_{\mu\nu}\ ;\qquad
\bar\chi=\chi+\eta\ .
\ee

For the limited purposes of this paper, quantum gravity means the theory of 
interacting fields $h_{\mu\nu}$ and $\eta$ 
propagating on the backgrounds $g_{\mu\nu}$ and $\chi$.
The infinitesimal form of diffeomorphism and Weyl transformations is
\bea
\delta_\xi\bar g_{\mu\nu}=&\calL_\xi \bar g_{\mu\nu}\ ;\qquad
\delta_\xi\bar\chi=\calL_\xi \bar\chi\ ,\\
\delta_\omega\bar g_{\mu\nu}=&2\omega\bar g_{\mu\nu}\ ;\qquad
\delta_\omega\bar\chi=-\omega\bar\chi\ ,
\eea
where $\xi$ and $\omega$ are infinitesimal transformation parameters
and $\calL_\xi$ is the Lie derivative along $\xi$.
There are two ways of splitting this infinitesimal transformation
between background and quantum parts:
the ``quantum gauge transformation'' is
\bea
\delta_\xi g_{\mu\nu}&=0\ ;
&\delta_\xi\chi=0\ ;
\\
\delta_\xi h_{\mu\nu}&=\calL_\xi \bar g_{\mu\nu}\ ;
&\delta_\xi\eta=\calL_\xi \bar\chi\ ,
\\
\delta_\omega g_{\mu\nu}&=0\ ;
&\delta_\omega\chi=0\ ,
\\
\delta_\omega h_{\mu\nu}&=2\omega\bar g_{\mu\nu}\ ;
&\delta_\omega\chi=-\omega\bar\chi\ ,
\eea
while the ``background gauge transformation'' is
\bea
\delta_\xi g_{\mu\nu}&=\calL_\xi g_{\mu\nu}\ ;\qquad
\delta_\xi\chi&=\calL_\xi \chi\ ;
\\
\delta_\xi h_{\mu\nu}&=\calL_\xi h_{\mu\nu}\ ;\qquad
\delta_\xi\eta&=\calL_\xi \eta\ ,
\\
\delta_\omega g_{\mu\nu}&=2\omega g_{\mu\nu}\ ;\qquad
\delta_\omega\chi&=-\omega\chi\ ,
\\
\delta_\omega h_{\mu\nu}&=2\omega h_{\mu\nu}\ ;\qquad
\delta_\omega\eta&=-\omega\eta\ ,
\eea

We choose background gauge conditions that break the quantum transformations,
as required to make the Hessian invertible,
but preserve invariance under the background transformations.
For diffeomorphisms we choose the gauge fixing action
\be
S_{GF} =  \frac{1}{2\alpha}\int d^4x 
\sqrt{g}\,\frac{1}{2}Z\chi^2 F_\mu {\bar g}^{\mu\nu} F_\nu\ ,
\ee
where
\be
\label{gf}
F_\nu =D_\mu h^\mu{}_\nu -\frac{\beta + 1}{4}D_\nu h\ .
\ee
Here $\alpha$ and $\beta$ are gauge parameters and $Z$ is a 
wave function renormalization constant to be specified later.
The ghost action corresponding to the gauge \eq{gf} is given by
\be
S_{gh}=\int d^4x \sqrt{g}\, \chi^2{\bar C}_\mu g^{\mu\nu} (\cO_{gh})_\nu^\rho C_\rho
=\ \cG_{gh}\left(\bar C,\cO_{gh} C\right)\ .
\label{gh}
\ee
where $\bar C$ and $C$ are dimensionless anticommuting vector fields,
$\cG_{gh}$ is the Weyl invariant inner product on vector fields
defined in \eq{wimM}, and
\be
\label{gravghost}
(\cO_{gh})_\mu^\nu=
-\frac{1}{\chi^2}\left(\delta_\mu^\nu D^2 +\frac{1-\beta}{2} D_\mu D^\nu +\cR_\mu{}^\nu\right)
\ee
is the Weyl-covariant operator acting on ghosts.
To gauge--fix Weyl invariance we impose that $\eta=0$, a condition that does not lead to ghosts.
With this condition we can simply delete from the Hessian the rows and columns
that involve the $\eta$ field and we remain with a Hessian that is a quadratic
form in the space of the covariant symmetric tensors $h_{\mu\nu}$.

In this space we choose the Weyl--invariant functional metric
\be
\cG_G(h,h')=
\int d^4x\sqrt{g}\chi^4 h_{\mu\nu}g^{\mu\rho}g^{\nu\sigma}h'_{\rho\sigma}\ .
\ee
which can be used to turn the Hessian into a linear operator $\cO_G$
acting on the space of covariant symmetric tensors.
Because the original action was Weyl--invariant,
this operator is Weyl-covariant, in the sense that
\begin{align}&
\label{henry}
(\cO_{G(\Omega^2 g_{\mu\nu},\Omega^{-1}\chi)})_{\mu\nu}{}^{\rho\sigma}(\Omega^2 h_{\rho\sigma})
=\Omega^2 (\cO_{G(g_{\mu\nu},\chi)})_{\mu\nu}{}^{\rho\sigma}h_{\rho\sigma}\ .
\end{align}
Likewise the operator \eq{gravghost} is Weyl-covariant.
From here there follows that the FRGE, which is a sum of traces of functions
of these operators, is Weyl--invariant.
This shows that there exists a quantization scheme which preserves Weyl--invariance along the flow,
so if one starts from a ``bare'' action which is Weyl invariant, the effective action
will also be Weyl--invariant.

Note that if we choose the background gauge such that $\chi=\mu$
(the background gauge being completely independent from the gauge fixing in the functional integral)
one obtains an effective action which is a functional of the background metric only.
This is exactly the same functional that one would have obtained
by integrating with the Weyl-non-invariant measure where $\chi$ is replaced by $\mu$
and with the action written in the same gauge.
Thus, also in the case of dynamical gravity, 
the choice of the gauge $\chi=\mu$ commutes with quantization.
By a similar argument one sees that the St\"uckelberg procedure of Weyl-covariantizing
an action also commutes with quantization.

\subsection{Example: one loop beta functions of Einstein gravity}

The above conclusion is completely general, but in order to
illustrate it with a concrete calculation we consider the simple case of
gravity with the Einstein-Hilbert action:
\be
\label{hilbert}
S(g) =   \frac{1}{16\pi G}\int d^4 x \sqrt{g}(2\Lambda-R) 
\ ,
\ee
Applying the Weyl covariantization (St\"uckelberg) procedure
described in section 3, this can be rewritten as
\be
\label{confaction}
S(g,\chi)=\int d^4 x \sqrt{g}\left[\lambda Z^2\chi^4
-\frac{1}{12}Z\chi^2\cR
\right]
\ , 
\ee
where
\be
\label{conversion}
\cR=R-6\chi^{-1}\Box\chi\ ;\qquad
Z\chi^2=\frac{12}{16\pi G}\ ;\qquad
\lambda=\frac{2\pi}{9}G\Lambda\ .
\ee

Neglecting interactions, the gravitons contribute to the one loop effective action
the terms 
\be
S(g,\chi)+\frac{1}{2}\Tr\log{\cO}_G-\Tr\log{\cO}_{gh}\ ,
\ee
which has to be added to the matter effective action.
For the explicit form of the operators we refer to \cite{percacci}.
Using the proper time representation \eq{propertime},
and the heat kernel expansion \eq{hk},
the effective action has quartic and quadratic divergences
which can be absorbed by redefining the bare couplings $Z_B$ and $\lambda_B$
that are present in the bare action $S$. In the gauge $\beta=\alpha=1$,
the renormalization conditions can be written in the form 
\begin{align}
\frac{1}{12}Z\chi^{2}=\frac{1}{12}Z_{B}\chi^{2}
-\frac{1}{6}\frac{1}{(4\pi)^{2}}\left(23+2n_{M}-n_{D}\right)\left(\Lambda_{UV}^2-k^2\right)\ ,
\\
\lambda Z^{2}\chi^{4}=\lambda_{B}Z_{B}^{2}\chi^{4}
-\frac{1}{8}\frac{1}{(4\pi)^{2}}\left(2+n_S+2n_{M}-4n_{D}\right)\left(\Lambda_{UV}^4-k^4\right)\ ,
\end{align}
where $k$ can be viewed here as a renormalization scale.
Observe that with these renormalization conditions the effective action
is the same as one would have obtained by cutting off
the $t$--integration in \eq{propertime} at $t=1/k^2$.
In other words, $k$ behaves exactly as an infrared cutoff.
These renormalization conditions may look a bit strange because 
of the appearance of the field $\chi$.
However, we assume here that both $\Lambda_{UV}$
and $k$ are constant multiples of the dilaton.
Defining $\hat\Lambda_{UV}=\Lambda_{UV}/\chi$, $u=k/\chi$ 
the renormalization conditions become
\begin{align}
\frac{1}{12}Z(u)=\frac{1}{12}Z_{B}(\hat{\Lambda}_{UV})-\frac{1}{6}\frac{1}{(4\pi)^{2}}\left(23+2n_{M}-n_{D}\right)\left(\hat{\Lambda}_{UV}^{2}-u^{2}\right)\ ,
\\
\lambda(u)Z^{2}(u)=\lambda_{B}(\hat{\Lambda}_{UV})Z_{B}^{2}(\hat{\Lambda}_{UV})-\frac{1}{8}\frac{1}{(4\pi)^{2}}\left(2+n_S+2n_{M}-4n_{D}\right)\left(\hat{\Lambda}_{UV}^{4}-u^{4}\right)\ .
\end{align}
Taking a $u$ derivative and adding the matter contributions, the full beta functions are 
\begin{align}
\label{fullbetas}
u\frac{dZ}{du}&=\frac{1}{4\pi^{2}}\left(23+2n_{M}-n_{D}\right)u^{2}
\\
u\frac{d\lambda}{du}&=\frac{2+n_S+2n_{M}-4n_{D}}{32\pi^{2}Z^{2}}u^{2}\left[u^{2}-16\lambda Z \frac{23+2n_{M}-n_{D}}{2+n_S+2n_M-4n_D}\right]
\end{align}
These beta functions agree (for the gravitational part) with those
computed in \cite{percacci} using the FRGE.
They depend explicitly on the independent variable $u$.
This is due to the fact that we are measuring
all dimensionful quantities in units of the dilaton.
If we measured dimensionful couplings in units of $k$,
the beta functions would not contain $k$ explicitly.
One can check that rewriting these equations for the variables
$\tilde\Lambda=\Lambda/k^2=6\lambda Z/u^2$ and $\tilde G=Gk^2=3u^2/4\pi Z$
one recovers the familiar beta functions of the Einstein--Hilbert truncation
\cite{reuter1,litim1,cpr2} and, in the presence of matter \cite{dou,perini1,largen}.

The equations (\ref{fullbetas}), together with the IR boundary conditions 
$Z(0)=Z_0,\lambda(0)=\lambda_0$, admit the general solution
\begin{align}
\label{rgsol}
Z(u)&=Z_0+\frac{23+2n_{M}-n_{D}}{8\pi^2}u^2\ ,
\\
\lambda(u)&=
\frac{\pi^2((2+n_S+2n_M-4n_D)u^4+128\pi^2Z_0^2\lambda_0)}{2(8\pi^2 Z_0+(23+2n_{M}-n_{D}) u^2)^2}\ .
\end{align}
The gravitational fixed point that one finds in the Einstein--Hilbert truncation
corresponds to the behavior for large $u$, where 
$\lambda(u)\to\lambda_{\star}
=\frac{\pi^2(2+n_S+2n_M-4n_D)}{2(23+2n_{M}-n_{D})^2}$
and $Z(u)\to Z_* u^2$ with $Z_*=\frac{23+2n_{M}-n_{D}}{8\pi^2}$. This is consistent with the notion of a fixed point
because the wave function renormalization $Z$ is a redundant coupling \cite{perini3}.

One can compute the one--loop contribution of gravitons to the effective action
by using non--local heat kernel techniques.
The first term in the curvature expansion is similar to the one
given in section 4.2.2 for a scalar field.

\section{Discussion}

In discussions of conformal invariance, misunderstandings frequently arise
due to the different physical interpretation of the transformations
that are used by different authors.
In particle physics language, a theory that contains dimensionful parameters is 
obviously not conformal.
Thus conformal invariance is a property of a very restricted class of theories.
In particular, in quantum field theory the definition of the
path integral generally requires the use of dimensionful parameters
(cutoffs, renormalization points) which break conformality
even if it was present in the original classical theory.
True conformality is only achieved at a fixed point of the renormalization group.
Let us call this the point of view I.

On the other hand in Weyl's geometry and its subsequent ramifications,
conformal (Weyl) transformations are usually interpreted as relating 
different local choices of units.
Since the choice of units is arbitrary and cannot affect the physics,
it follows that essentially any physical theory can be formulated in a
Weyl--invariant way.
This point of view is more common among relativists.
Let us call it the point of view II.

The way in which a generic theory containing dimensionful parameters
can be made Weyl--invariant is by allowing those parameters to become
functions on spacetime, {\it i.e.} to become fields.
This is the step that the adherents of the interpretation I are generally
unwilling to make, since then one would have to ask
whether these fields have a dynamics of their own or not, and, in the quantum case,
whether they have to be functionally integrated over or not.
It can be unnatural to have fields in the theory that do not obey some
specific dynamical equation, and it is clear that in general, if one allows all the
dimensionful couplings to become dynamical fields, the theory is physically
distinct from the original one.

There is however one way in which Weyl--invariance can be introduced in any theory
without altering its physical content, and that is to introduce {\it a single}
scalar field, which we called a dilaton 
(sometimes also called a ``St\"uckelberg'' or ``Weyl compensator'' or ``spurion'' field)
and to assume that all dimensionful parameters are proportional to it.
This field carries a nonlinear realization of the Weyl group,
since it is not allowed to become zero anywhere.
Even though the new field obeys dynamical equations, it does not modify
the physical content of the theory because it
is exactly neutralized by the enlarged gauge invariance. 
In practice, it can be eliminated by choosing the Weyl gauge
such that it becomes constant.

All this is well--known in the classical case. 
It had already been observed both in a perturbative and nonperturbative context that
the above considerations can be generalized to the context of quantum field theory
by treating the cutoff or the renormalization point in the same way
as the mass or dimensionful parameters that are present in the action.
In this paper we have discussed in particular the formulatiom
of the renormalization group using the point of view II.
It has proven convenient to adopt a non--perturbative definition of the
renormalization group, where one considers the dependence of the effective action on an
externally prescribed smooth cutoff $k$.
The advantage of this procedure is that the resulting ``beta functional'' is both
UV and IR finite and one can use it to define a first order differential equation
whose solution, for $k\to0$, is the effective action.
It can therefore be viewed as a non--perturbative way of defining
(and calculating) the effective action.
Using this method we have shown in complete generality that one can
define a flow of Weyl--invariant actions whose IR endpoint is 
a Weyl--invariant effective action.
This is our main result.

This provides an answer to the following question.
Suppose we start from a theory that contains dimensionless parameters,
and recast it in a Weyl--invariant form by introducing a dilaton field.
If we quantize this Weyl--invariant theory, is the result equivalent to
the one we would have obtained by quantizing the original theory?
The answer is affirmative, if we use throughout ({\it i.e.} for all fields)
the Weyl--invariant measure.
\footnote{By contrast, suppose that after having quantized the
matter fields we also quantize the metric and dilaton, using the standard,
Weyl--non--invariant measure I.
(One does not need to have a full quantum gravity for this argument, 
it is enough to think of a one loop calculation in the context 
of an effective field theory).
The integration over metric and dilaton will now proceed with total 
actions $S_G+\GammaI$ and $S_G+\GammaII$,
depending on whether we used for the matter the measures I or II.
Clearly the resulting theories are physically inequivalent:
In the first case the action is not Weyl invariant, so the
dilaton field is physical, in the second case the action is Weyl invariant
and the dilaton can be gauged away.
So, {\it all else being equal}, quantizing matter with measures I and II
leads to physically different theories.}
Thus, there is a quantization procedure that commutes with the St\" uckelberg trick.

It is important to understand that although Weyl invariance is not anomalous,
there is still a trace anomaly, in the sense that the trace of the energy--momentum,
which is classically zero, is not zero in the quantum theory.
This can be easily understood from the fact that in the Weyl--invariant quantization
one obtains an effective action that depends not only on the metric but also
on the dilaton. Weyl--invariance of the effective action is compatible with
a nonvanishing trace, because the latter cancels out against the variation of the dilaton. 
(We have provided fully explicit examples of this phenomenon in section 4.2.1.)
The physics of the Weyl--invariant quantization procedure is completely equivalent to the standard one. In particular, all the proposed physical applications of the
trace anomaly remain valid \cite{fulling,mottola,urban}.

In order not to modify the dynamics we have used the ``only one dilaton'' prescription,
with the consequence that the cutoff and the dilaton are proportional.
The proportionality factor, which we have called $u$, is the RG parameter
in this formulation. It is dimensionless (since it expresses the cutoff
relative to the unit of mass), Weyl--invariant and constant on spacetime.
Thus, running couplings are functions $g(u)$.
In this we differ from other approaches to dilaton dynamics where the
couplings depend on a mass scale $k$ that is allowed to be a function on spacetime.
In practice the difference is not so important, because the variation
of $g(u)$ with respect to $k$, keeping $\chi$ constant, is the same that one would
obtain if one assumed that $g$ is a function of $k$.

Given that in this formalism all theories are conformally invariant,
one can also ask what is special about conformal field theories
(in the standard sense of quantum field theory),
and in particular about fixed points of the renormalization group.
The answer is that for generic theories, conformal invariance is
only achieved at the price of having a dilaton in the effective action.
True conformal field theories are conformal even without the dilaton,
so one must expect that as the RG flow approaches a fixed point, the dilaton must decouple.

Weyl--invariance is the statement of conformal invariance in a 
general relativistic setting, so we expect the formalism developed here to
be especially relevant in the discussion of a possible fixed point for gravity.
Barring exotic phenomena, one would expect that a gravitational fixed point must
correspond to a Weyl--invariant, as opposed to merely globally scale invariant--theory.
We have illustrated in section 6.1 the appearance of a gravitational fixed point
in the case of the Einstein--Hilbert truncation. It would be interesting to
extend the discussion to higher derivative terms.
In particular it has been observed in \cite{benedetti} that in an $f(R)$--truncation
a fixed point would correspond to an effective action that is proportional to $R^2$.
As such, this would only be globally scale--invariant. We conjecture that 
the fixed point is conformal and that this term should
be interpreted as a piece of the Euler term, which is the only local, Weyl--invariant
combination of curvatures besides the square of the Weyl tensor.
Consistently with this, we observe that the RG trajectory
that corresponds to the fixed point discussed in section 6.1,
corresponds to putting $Z_0=0$ in \eq{rgsol}.
The corresponding EAA is given by \eq{confaction} with $Z(u)=Z_*u^2$ and $\lambda(u)=\lambda_*$.
When we take the $u\to0$ limit, to obtain the effective action, we find zero.

We conclude by mentioning some possible extensions and applications of this work.
Recalling that here we have analyzed the case of integrable Weyl geometry,
it is natural to extend our results to the non--integrable case. This work is in progress.
The example in section 6.2 deals with a truncation of the gravitational action
to terms involving at most two derivatives.
In four dimensions, such terms necessarily have dimensionful couplings.
In view of the remarks in the preceding paragraph, it seems interesting to
re--analyze the terms with four derivatives, in particular the Weyl--squared
and the Euler term.
Perhaps the most important applications of Weyl geometry is to cosmology,
where it is often useful to change conformal frame.
Even at a classical level, there has been some controversy on the issue
whether such frames should be interpreted as
defining physically equivalent situations. 
Our point of view agrees with that of \cite{flanagan}.
The question is much more delicate in the quantum theory, however.
For explicit quantum calculations where different conformal frames
can be seen to yield equivalent physics, see \cite{shapo2}.
The present work provides a general proof that
with a suitable quantization procedure, the equivalence between conformal frames
can be maintained also in the quantum theory.
One could use this to study the relation between 
$f(R)$ and scalar--tensor theories at the quantum level.

\bigskip
\bigskip
\goodbreak


\begin{thebibliography}{99}


\bibitem{duff}
D.M. Capper and M.J. Duff, 
Nuovo Cim. A {\bf 23}, 173 (1974);
Phys. Lett. A {\bf 53}, 361 (1975);\\
M.J. Duff, 
Nucl. Phys. B {\bf 125}, 334 (1977);
Class. and Quantum Grav. {\bf 11}, 1387 (1994)
[arXiv: hep-th/9308075].

\bibitem{englert}
F. Englert, C. Truffin and R. Gastmans,
Nucl. Phys.B  {\bf 117}, 407 (1976).

\bibitem{fv}
E.S. Fradkin, G.A. Vilkovilsky, Phys. Lett. B {\bf 73}, 209 (1978).

\bibitem{flop}
R.Floreanini, R. Percacci,
Nucl. Phys. B {\bf 436}, 141 (1995)
[arXiv: hep-th/9305172];
Phys. Rev. D {\bf 52}, 896 (1995)
[arXiv: hep-th/9412181].

\bibitem{reuterliouville}
M. Reuter, C. Wetterich,
Nucl. Phys. B {\bf 506}, 483 (1997) 
[arXiv: hep-th/9605039];
\\
M. Reuter, 
[arXiv: hep-th/9612158].

\bibitem{creh}
M. Reuter, H. Weyer,
Phys. Rev. D {\bf 79}, 105005 (2009) [arXiv: 0801.3287];
Phys. Rev. D {\bf 80}, 025001 (2009) [arXiv: 0804.1475].

\bibitem{machado}
P.F. Machado, R. Percacci,
Phys. Rev. D {\bf 80}, 024020 (2009) [arXiv: 0904.2510].

\bibitem{shapo}
M. Shaposhnikov, D. Zenhausern,
Phys. Lett. B {\bf 671}, 162 (2009) [arXiv: 1104.1392];\\
M. Shaposhnikov, I. Tkachev,
Phys. Lett. B {\bf 675}, 403 (2009) [arXiv: 0811.1967].

\bibitem{iorio}
A. Iorio, L. O'Raifeartaigh, I. Sachs, C. Wiesendanger
Nucl. Phys. B {\bf 495}, 433 (1997).

\bibitem{jackiw}
R. Jackiw, S.-Y. Pi,
J. Phys. A {\bf 44} (2011) 223001 [arXiv: 1101.4886];
\\
Sheer El-Showk, Yu Nakayama, Slava Rychkov
Nucl. Phys. B {\bf 848}, 578 (2011) [arXiv: 1101.5385].


\bibitem{fortin}
J.F. Fortin, B. Grinstein, A. Stergiou 
Phys. Lett. B {\bf 704}, 74 (2011) [arXiv: 1106.2540];
JHEP {\bf 1207}, 025 (2012) [arXiv:1107.3840].

\bibitem{hawking}
S. Hawking,
Commun. Math. Phys. {\bf 55}, 133 (1977). 

\bibitem{raja}
R. Jackiw, R. Rajaraman,
Phys. Rev. Lett. {\bf 54}, 1219 (1985); {\it ibid.} {\bf 54}, 2060 (1985);
\\
P. Mitra, R. Rajaraman,
Phys. Lett. B {\bf 225}, 267 (1989);
\\
R. Percacci, R. Rajaraman,
Int. J. Mod. Phys. A {\bf 4}, 4177 (1989).

\bibitem{wetterich}
C. Wetterich,
Phys. Lett. B {\bf 301}, 90 (1993).
T. Morris,
Phys. Lett. B {\bf 329}, 241 (1994) [arXiv: hep-ph/9403340].

\bibitem{reuter1}
M. Reuter,
Phys. Rev. D {\bf 57}, 971 (1998) [arXiv: hep-th/9605030].

\bibitem{polyakov}
A. Polyakov, 
Phys. Lett. B {\bf 103}, 207 (1981).

\bibitem{codello1}
A. Codello, 
Ann. Phys. {\bf 325}, 1727 (2010) [arXiv: 1004.2171] 

\bibitem{russian1}
A.O. Barvinsky, G.A. Vilkovisky,
Nucl. Phys. B {\bf 333}, 471 (1990).

\bibitem{coza}
A. Codello, O. Zanusso, [arXiv: 1203.2034].

\bibitem{codello2}
A. Codello, 
New J. Phys. {\bf 14}, 015009 (2012) [arXiv: 1108.1908].

\bibitem{deser}
S. Deser and A. Schwimmer,
Phys. Lett. B {\bf 309}, 279 (1993) [arXiv: hep-th/9302047];
\\
S. Deser, Phys. Lett. B {\bf 479}, 315 (2000).

\bibitem{erdmenger}
J. Erdmenger, Class. Quant. Grav. {\bf 14}, 2061 (1997);
\\
J. Erdmenger and H. Osborn, Nucl. Phys. B {\bf 483}, 431 (1997) [arXiv: hep-th/9605009].

\bibitem{mottola flat}
C. Coriano, L. Delle Rose, E. Mottola, M. Serino, 
JHEP {\bf 1208}, 147 (2012) [arXiv: 1203.1339]

\bibitem{russian2}
A. O. Barvinsky, Yu. V. Gusev, G. A. Vilkovisky, V. V. Zhytnikov, 
J. Math. Phys. {\bf 35}, 3525 (1994), [arXiv: gr-qc/9404061];
{\it ibid.} {\bf 35}, 3543 (1994) [arXiv: gr-qc/9404063];
A. O. Barvinsky, Yu. V. Gusev, V. V. Zhytnikov,  G. A. Vilkovisky,
[arXiv: 0911.1168]. 

\bibitem{russian3}
A.O. Barvinsky, Yu.V. Gusev, G.A. Vilkovisky, V.V. Zhytnikov,
Nucl. Phys. B {\bf 439}, 561 (1995); [arXiv: hep-th/9404187].

\bibitem{riegert}
R.J. Riegert, Phys. Lett. B {\bf 134}, 56 (1984).

\bibitem{mottola1}
P. O. Mazur, E. Mottola, Phys. Rev. D {\bf 64}, 104022 (2001).

\bibitem{russian4}
A.O. Barvinsky, A.G. Mirzabekian, V.V. Zhytnikov, 
talk given at Conference: C95-06-12.3 [arXiv: gr-qc/9510037].

\bibitem{hr}
U. Harst, M. Reuter,
JHEP {\bf 1205}, 005 (2012) [arXiv: 1203.2158].

\bibitem{dp}
P. Dona', R. Percacci, [arXiv: 1209.3649].

\bibitem{background}
M. Reuter, H. Weyer,
Gen. Rel. Grav. {\bf 41}, 983 (2009) [arXiv: 0903.2971].

\bibitem{percacci}
R. Percacci, 
New J. Phys. {\bf 13}, 125013 (2011) [arXiv: 1110.6758].

\bibitem{litim1}
D.F. Litim,
Phys. Rev. Lett. {\bf 92}, 201301 (2004)
[arXiv: hep-th/0312114].

\bibitem{cpr2}
A. Codello, R. Percacci, C. Rahmede,
Annals Phys. {\bf 324}, 414 (2009)
[arXiv: 0805.2909].

\bibitem{dou}
D. Dou and R. Percacci,
Class. Quant. Grav.  {\bf 15}, 3449 (1998) [arXiv: hep-th/9707239].

\bibitem{perini1}
R. Percacci and D. Perini,
Phys. Rev. D {\bf 67}, 081503 (2003) [arXiv: hep-th/0207033];
Phys. Rev. D {\bf 68}, 044018 (2003) [arXiv: hep-th/0304222].

\bibitem{largen}
R. Percacci,
Phys. Rev. D {\bf 73}, 041501 (2006) [arXiv: hep-th/0511177].

\bibitem{perini3}
R. Percacci and D. Perini, Class. Quant. Grav. {\bf 21}, 5035 (2004)
[arXiv: hep-th/0401071]; 
R. Percacci, J. Phys. A {\bf 40}, 4895 (2007)
[arXiv: hep-th/0409199].

\bibitem{fulling}
S.M. Christensen, S.A. Fulling, 
Phys. Rev. D {\bf 15}, 2088 (1977).

\bibitem{urban}
F.R. Urban, A.R. Zhitnitsky,
Nucl. Phys. B {\bf 835}, 135 (2010) [arXiv: 0909.2684];
\\ 
E.C. Thomas, F.R. Urban, A.R. Zhitnitsky,
JHEP {\bf 0908}, 043 (2009) [arXiv: 0904.3779]

\bibitem{mottola}
E. Mottola, J. Phys. Conf. Ser. {\bf 314}, 012010 (2011) 
[arXiv: 1107.5086];
E. Mottola and R. Vaulin, Phys. Rev. D {\bf 74}, 064004 (2006) 
[arXiv: gr-qc/0604051]

\bibitem{benedetti}
D. Benedetti, F. Caravelli,
JHEP {\bf 1206}, 017 (2012) [arXiv: 1204.3541].

\bibitem{flanagan}
E.E. Flanagan, 
Class. and Quantum Grav. {\bf 21}, 3817 (2004) 
[arXiv: gr-qc/0403063].

\bibitem{shapo2}
F. Bezrukov, A. Magnin, M. Shaposhnikov, S. Sibiryakov,
JHEP {\bf 1101}, 016 (2011) [arXiv: 1008.5157];
\\ 
F. Bezrukov, D. Gorbunov, M. Shaposhnikov,
JCAP {\bf 1110}, 001 (2011) [arXiv: 1106.5019].

\end{thebibliography}
\end{document}